\documentclass[10pt]{article}
\usepackage{amsmath,amssymb}
\usepackage[pdftex]{graphicx}
\usepackage[hidelinks]{hyperref}
\usepackage{tikz}
\usetikzlibrary{intersections, calc, arrows.meta, cd}
\usepackage{amsthm}
\usepackage{extarrows}
\usepackage{txfonts}
\usepackage{here}
\usepackage{genyoungtabtikz} %%ヤング図形
\usepackage{authblk}

\theoremstyle{definition}
\newtheorem{theorem}{Theorem}

\newtheorem{lemma}[theorem]{Lemma}

\newtheorem{example}[theorem]{Example}

\numberwithin{theorem}{section}
\numberwithin{equation}{section}
\numberwithin{rem}{theorem}

\newcommand{\BS}[1]{\boldsymbol{#1}}
\newcommand{\rbin}{\mathop{\mathrm{bin}}}
\newcommand{\lbin}{{\mathop{\mathrm{bin}}}^\prime}

\newcommand{\relmiddle}[1]{\mathrel{}\middle#1\mathrel{}}

\date{}

\begin{document}
  \title{Linearization of the box-ball system with box capacity $L$}
  \author[$\dagger$]{Atsushi Maeno}
  \author[$\dagger$]{Satoshi Tsujimoto}
  \affil[$\dagger$]{Department of Applied Mathematics and Physics, Graduate School of Informatics,\par  Kyoto University, Kyoto, 606-8501, Japan}
 \maketitle
 \begin{abstract}
  We construct a bijection between the state of the box-ball system with box capacity $L$ and a pair of two sequences. During the time evolution, one of the sequences moves at speed 1, and the other follows the rules of the box-ball system with box capacity one, which can be linearized by the Kerov-Kirillov-Reshetikhin(KKR) bijection. Our method can be applied to a state including a negative value or a value greater than the box capacity.
 \end{abstract}
 %\tableofcontents
 
\section{Introduction}\label{sec1}
  In 1990, Takahashi and Satsuma introduced a soliton cellular automaton called the box-ball system (BBS) \cite{TS}. The state of the original BBS is described with infinitely many boxes and finitely many balls, where each box can hold at most one ball (that is, box capacity is one). It has since been studied from various perspectives, including ultradiscretization of soliton equations \cite{TH,TTMS}, crystal bases \cite{KOSTY}, and the inverse scattering method \cite{Ta1,WRSG}.
  
  Kuniba et al. found that the time evolution of the BBS can be linearized using the Kerov-Kirillov-Reshetikhin (KKR) bijection \cite{KOSTY,Ta1}. The KKR bijection was originally introduced for the analysis of solvable lattice models \cite{KKR} and was later investigated in relation to Kashiwara crystals. For the BBS with box capacity one, the procedure to compose the KKR bijection was simplified with ``01-arc lines''\cite{KS}, and an elementary proof was given by Kakei et al. \cite{KNTW} The 01-arc lines are used to describe the time evolution of the BBS, and they are applied to BBS analysis in the context of probability theory \cite{CKST}.

  In this paper, we give a method of linearizing the time evolution of the BBS with box capacity $L$ by decomposing a state into two sequences: 1) a sequence that shifts to the right at speed 1 and 2) a binary sequence that exhibits the time evolution of the BBS with box capacity. Our decomposition method can be easily applied to a state with a negative value or a value greater than the box capacity.

  We use the following notation in this paper: 
  \begin{itemize}
   \item Semi-infinite integer sequence: $\BS{\eta}=(\eta_0, \eta_1, \eta_2, \ldots), \eta_j\in\mathbb{Z}\ (j=0,1,2,\ldots)$.
   \item The $j$-th component of $\BS{\eta}$: $(\BS{\eta})_j=\eta_j.$
  \end{itemize}

 \section{\texorpdfstring{BBS with Box Capacity $L$ for a Sequence of Integer Values}{}}\label{sec2}
  \subsection{Time evolution}\label{sec2.1}
   In this paper, we consider the BBS with box capacity $L\in\mathbb{Z}_{>0}$ (hereinafter, we call this BBS($L$)). First, we introduce the original BBS($L$) for $L+1$ values. The set of BBS($L$) states is denoted by $\mathcal{S}_L$: 
   \begin{align}\label{eq2.1}
    \mathcal{S}_L &= \left\{\BS{\eta}\in\{0,1,\ldots,L\}^{\mathbb{Z}_{\geq0}}\relmiddle| \sum_{j=0}^\infty \eta_j<\infty, \eta_0 = 0, \sum_{j=0}^i(L-\eta_j)\geq \sum_{j=0}^{i+1}\eta_j \,(i=0, 1, \ldots)\right\}.
   \end{align}

   The time evolution $T_L: \mathcal{S}_L\to\mathcal{S}_L;\BS\eta^t\mapsto\BS\eta^{t+1}=T_L(\BS\eta^t)$ of the BBS($L$) can be described with a carrier that transports balls from left to right according to the following rules: 
   \begin{enumerate}\renewcommand{\labelenumi}{(\roman{enumi})}
    \item The carrier starts from the leftmost site with no balls, and it runs to the right.
    \item When the carrier passes in front of the $j$-th box, it performs the following two operations simultaneously: 
     \begin{itemize}
      \item if that box contains at least one ball, the carrier picks the ball(s) up,
      \item if that box is not full and the carrier has at least one ball, the carrier drops off as many balls as possible into that box.
     \end{itemize}
    \item When all the balls have been transported to another box, the carrier stops.
   \end{enumerate}
   By repeating this procedure, the time evolution series of BBS($L$) can be obtained (see Example \ref{ex2.1}).

   The time evolution $T_L$ can be rewritten as a piecewise linear equation known as the ultradiscrete Korteweg-de Vries(uKdV) equation: 
   \begin{align}
    u_0^t &= 0, \label{eq2.2}\\
    \eta_j^{t+1} &= \min(L-\eta_j^t, u_j^t),\label{eq2.3}\\
    u_{j+1}^t &= \eta_j^t+u_j^t-\eta_j^{t+1}\nonumber\\
     &= \eta_j^t + \max(0, \eta_j^t+u_j^t-L),\label{eq2.4}
   \end{align}
   where $\eta_j^t$ is the number of balls in the $j$-th box at time $t$, and $u_j^t$ is the number of balls in the carrier just before passing the $j$-th box at time $t$. Eq. (\ref{eq2.4}) means that the total number of balls is conserved in the time evolution as $u_j^t+\eta_j^t = \eta_j^{t+1}+u_{j+1}^t$. 

   The BBS with finitely many balls has a uniquely determined reverse time evolution. Let $x$ be a non-negative integer such that $\eta_j^t=0$ for all $j\geq x$. Then, the variables satisfy
   \begin{align}
    u_x^{t-1} &= 0,\label{eq2.5}\\
    \eta_j^{t-1} &= \min(L-\eta_j^t, u_{j+1}^{t-1}),\label{eq2.6}\\
    u_j^{t-1} &= \eta_j^t+u_{j+1}^{t-1}-\eta_j^{t-1}\nonumber\\
     &= \eta_j^t + \max(0, \eta_j^t+u_{j+1}^{t-1}-L)\label{eq2.7}.
   \end{align}

   \begin{example}\label{ex2.1}
    Box capacity $L=3$.
    \begin{align*}
     \BS{\eta}^0 &= 0002320002101000000000000\cdots\\
     T_{\BS{L}}(\BS{\eta}^0) = \BS{\eta}^1 &= 0000013300210100000000000\cdots\\
     (T_{\BS{L}})^2(\BS{\eta}^0) = \BS{\eta}^2 &= 0000000033121010000000000\cdots\\
     (T_{\BS{L}})^3(\BS{\eta}^0) = \BS{\eta}^3 &= 0000000000212321000000000\cdots\\
     (T_{\BS{L}})^4(\BS{\eta}^0) = \BS{\eta}^4 &= 0000000000021012320000000\cdots\\
     (T_{\BS{L}})^5(\BS{\eta}^0) = \BS{\eta}^5 &= 0000000000002101013300000\cdots\\
     (T_{\BS{L}})^6(\BS{\eta}^0) = \BS{\eta}^6 &= 0000000000000210100033100\cdots
    \end{align*}
   \end{example}
   
   In the argument in the next section, we allow the state value $\eta_j^t$ to take a negative value or a value greater than the box capacity $L$. For a state of arbitrary integer values, let $M$ be the non-negative integer which corresponds to the maximal excess below 0 or above $L$, as $M = \max_{j\in\mathbb{Z}_{\geq0}}\max(-\eta_j^t, \eta_j^t-L)$. One can find that $M$ is invariant under the time evolution Eqs. (\ref{eq2.2}), (\ref{eq2.3}) and (\ref{eq2.4}). Define the set of states as

   \begin{align}
    \mathcal{S}_{L, M} &= \left\{\BS{\eta}\in\{-M, -M+1, \ldots, L+M\}^{\mathbb{Z}_{\geq0}} \mathrel{}\middle|\mathrel{} \sum_{j=0}^\infty |\eta_j|<\infty, \eta_0\leq 0, \right. \nonumber\\
    &\qquad\qquad \left.\max_{j\in\mathbb{Z}_{\geq0}}\max(-\eta_j^t, \eta_j^t-L)=M, \sum_{j=0}^{i}(L-\eta_j)\geq\sum_{j=0}^{i+1}\eta_j\, (i=0, 1, \ldots)\right\}. \label{eq2.8}
   \end{align}

   We set variables $\tilde{\eta}_j^t=\eta_j^t+M, \tilde{u}_j^t=u_j^t+M, \tilde{L}=L+2M$\cite{GNN2}. Then, the time evolution Eqs. (\ref{eq2.2}), (\ref{eq2.3}) and (\ref{eq2.4}) become

   \begin{align}
    \tilde{u}_0^t &= M, \label{eq2.9}\\
    \tilde{\eta}_j^{t+1} &= \min(\tilde{L}-\tilde{\eta}_j^t, \tilde{u}_j^t), \label{eq2.10}\\
    \tilde{u}_{j+1}^t &= \tilde{\eta}_j^t + \tilde{u}_j^t - \tilde{\eta}_j^{t+1}\nonumber\\
     &= \tilde{\eta}_j^t+\max(0, \tilde{\eta}_j^t+\tilde{u}_j^t-\tilde{L}).\label{eq2.11}
   \end{align}
   
   Since Eqs. (\ref{eq2.10}) and (\ref{eq2.11}) are equivalent to Eqs. (\ref{eq2.3}) and (\ref{eq2.4}) respectively, we can regard this system as a BBS($\tilde{L}$). Note that the carrier starts with $M$ balls and stops when the carrier and every box to the right of the carrier have $M$ balls. Let $T_L^M$ denote the time evolution defined by Eqs. (\ref{eq2.3}), (\ref{eq2.4}) with the initial value of the carrier $u_0=M$.

   The BBS($L$) can be expressed in terms of the BBS($1$) by considering a transformation between a binary sequence and an $L+1$ value sequence\cite{TS2}. We define two binary sequences $\BS{r}_I$ and $\BS{l}_{I^\prime}$ from a state $\BS\eta^t\in\mathcal{S}_{L, M}$. Let $J(\BS\eta^t)$ be a set of indices $j$, where $\eta_j^t+u_j^t\geq L$, 
   \begin{align}
    J(\BS\eta^t) &= \{j\in\mathbb{Z}_{\geq0}\mid \eta_j^t+u_j^t\geq L\}\nonumber\\
     &= \{j\in\mathbb{Z}_{\geq0}\mid \tilde{\eta}_j^t+\tilde{u}_j^t\geq\tilde{L}\}.\label{eq2.12}
   \end{align}
   For a subset $I=\{i_0, i_1, \ldots, i_m\}\subset J(\BS\eta^t)$, let $r_I(\tilde\eta_j)$ be the binary sequence of length $\tilde{L}$ as
   \begin{align}
    r_I(\tilde{\eta}_j^t) &= \begin{cases}
    1^{\tilde{\eta}_j^t}0^{\tilde{L}-\tilde{\eta}_j^t} & (j\in I),\\
    0^{\tilde{L}-\tilde{\eta}_j^t}1^{\tilde{\eta}_j^t} & (j\not\in I),
    \end{cases}\label{eq2.13}
   \end{align}
   and let $\BS{r}_I(\BS{\eta}^t)$ be the concatenation of $r_I(\tilde{\eta}_j^t)$
   \begin{align}
    \BS{r}_I(\BS\eta^t) = r_I(\tilde{\eta}_0^t)\cdot r_I(\tilde{\eta}_1^t)\cdot r_I(\tilde{\eta}_2^t)\cdots.\label{eq2.14}
   \end{align}

   Similarly, let $J^\prime(\BS\eta^t)$ be a set of indices $j$, where $\eta_j^t+u_{j+1}^{t-1}\geq L$, 
   \begin{align}
    J^\prime(\BS\eta^t) &= \{j\in\mathbb{Z}_{\geq0}\mid\eta_j^t+u_{j+1}^{t-1}\geq L\}\nonumber\\
     &= \{j\in\mathbb{Z}_{\geq0}\mid\tilde{\eta}_j^t+\tilde{u}_{j+1}^{t-1}\geq \tilde{L}\}.\label{eq2.15}
   \end{align}

   For a subset $I^\prime=\{i_0^\prime, i_1^\prime, \ldots, i_m^\prime\}\subset J^\prime(\BS\eta^t)$, let $l_{I^\prime}(\tilde{\eta}_j^t)$ be the binary sequence of length $\tilde{L}$ as
   \begin{align}
    l_{I^\prime}(\tilde{\eta}_j^t) &= \begin{cases}
    0^{\tilde{L}-\tilde{\eta}_j^t}1^{\tilde{\eta}_j^t} & (j\in I^\prime),\\
    1^{\tilde{\eta}_j^t}0^{\tilde{L}-\tilde{\eta}_j^t} & (j\not\in I^\prime),
    \end{cases}\label{eq2.16}
   \end{align}
   and let $\BS{l}_{I^\prime}(\BS\eta^t)$ be the concatenation of $l_{I^\prime}(\tilde{\eta}_j^t)$
   \begin{align}
    \BS{l}_{I^\prime}(\BS\eta^t) = l_{I^\prime}(\tilde{\eta}_0^t)\cdot l_{I^\prime}(\tilde{\eta}_1^t)\cdot l_{I^\prime}(\tilde{\eta}_2^t)\cdots.\label{eq2.17}
   \end{align}
 
   \begin{example}\label{ex2.3}$L=2$, $\BS\eta^t=0, 3, 0, -1, 3, 0, 2, -1, 0, 0, \ldots$.

    $M=\max_{j\in\mathbb{Z}_{\geq0}}\max(-\eta_j^t, \eta_j^t-L)=1$, $\tilde{L}=4$, and $\tilde{\BS\eta}^t=1, 4, 1, 0, 4, 1, 3, 0, 1, 1, \ldots$. Using time evolution Eqs. (\ref{eq2.2}), (\ref{eq2.3}) and (\ref{eq2.4}), we get
    \begin{align*}
     \BS{u}^t &= 0, 0, 4, 2, -1, 3, 1, 3, -1, 0, \ldots,\\
     \BS\eta^{t+1} &= 0, -1, 2, 2, -1, 2, 0, 3, -1, 0,\ldots,\\
     \tilde{\BS{u}}^t &= 1, 1, 5, 3, 0, 4, 2, 4, 0, 1, \ldots,\\
     \tilde{\BS\eta}^{t+1} &= 1, 0, 3, 3, 0, 3, 1, 4, 0, 1,\ldots.
    \end{align*}
    Then, $J(\BS\eta^t)=\{j\in\mathbb{Z}_{\geq0}\mid\tilde{\eta}_j^t+\tilde{u}_j^t\geq \tilde{L}\}$ becomes
    \begin{align*}
     J(\BS\eta^t) = \{1, 2, 4, 5, 6, 7\}.
    \end{align*}
    Choosing a subset of $J(\BS\eta^t)$ as $I=\{2, 5\}$, we obtain
    \begin{align*}
     \BS{r}_I(\BS\eta^t) &= 000111111000000011111000011100000001\cdots,\\
     T_1^1(\BS{r}_I(\BS\eta^t)) &= 100000000111111000000111100011110000\cdots,\\
     \BS{l}_I(T_{\BS{L}}(\BS\eta^t)) &= 100000000111111000000111100011110000\cdots.
    \end{align*}
   \end{example}

   \begin{theorem}\label{th2.2}
    For $\BS\eta^t\in\mathcal{S}_{L, M}$ and $I=\{i_0, i_1, \ldots, i_m\}\subset J(\BS\eta^t)$, 
    \begin{align}
    T_1^M(\BS{r}_I(\BS{\eta^t})) &= \BS{l}_I(T_L(\BS{\eta}^t)).\label{eq2.18}
    \end{align}
    \begin{proof}
     From Eq. (\ref{eq2.4}),  we have $\eta_j^t+u_j^t = \eta_j^{t+1}+u_{j+1}^t$ and $J(\BS\eta^t)=J^\prime(T_L(\BS\eta^t))$. Consider the carrier passing through $\tilde{L}$ boxes from the $j\tilde{L}$-th to the $(j+1)\tilde{L}-1$-th position in $\BS{r}_I(\BS\eta^t)$. 
     \begin{itemize}
      \item Case 1: $j\not\in I$ and $\tilde{\eta}_j^t+\tilde{u}_j^t<\tilde{L}$

       The carrier drops off $\tilde{u}_j^t$ balls into the first $\tilde{u}_j^t$ empty boxes and picks up $\tilde{\eta}_j^t$ balls as shown in Fig. \ref{fig1}. Then, the number of balls in the boxes in this interval becomes $\tilde{u}_j^t$, which is equal to $\tilde{\eta}_j^{t+1}=\min(\tilde{L}-\tilde{\eta}_j^t, \tilde{u}_j^t)$.
       %\begin{figure}[H]\centering
       \begin{figure}[htbp]\centering
        % \begin{tikzpicture}
        %  \draw[->, >=stealth, very thick] (-1.5, 0)--(1.5, 0);
        %  \draw[->, >=stealth, very thick] (0, 0.7)--(0, -0.7);
        %  \draw (-1.5, 0)node[left]{Carrier: $\tilde{u}_j^t$};
        %  \draw (1.5, 0)node[right]{$\tilde{u}_{j+1}^t$};
        %  \draw (0, 0.7)node[above]{$\underbrace{0\cdots\cdots0}_{\scalebox{1}{$\tilde{L}-\tilde{\eta}_j^t$}}\,\,\underbrace{1\cdots1}_{\scalebox{1}{$\tilde{\eta}_j^t$}}$};
        %  \draw (0, -0.7)node[below]{$\underbrace{1\cdots1}_{\scalebox{1}{$\tilde{\eta}_j^{t+1}=\tilde{u}_j^t$}}\underbrace{0\cdots\cdots0}_{\scalebox{1}{$\tilde{L}-\tilde{u}_j^t$}}$};
        % \end{tikzpicture}
        \includegraphics[height=3cm]{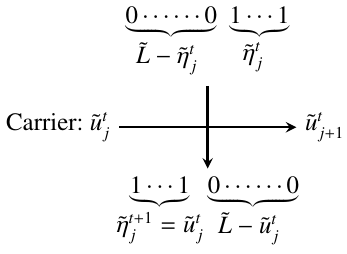}
        \caption{\label{fig1}Case 1 in Theorem \ref{th2.2}.}
       \end{figure}

      \item Case 2: $j\not\in I$ and $\tilde{\eta}_j^t+\tilde{u}_j^t\geq \tilde{L}$

       The carrier drops off $\tilde{L}-\tilde{\eta}_j^t$ balls into $\tilde{L}-\tilde{\eta}_j^t$ empty boxes and picks up $\tilde{\eta}_j^t$ balls as shown in Fig. \ref{fig2}. Then, the number of balls in the boxes in this interval becomes $\tilde{L}-\tilde{\eta}_j^t$, which is equal to $\tilde{\eta}_j^{t+1}=\min(\tilde{L}-\tilde{\eta}_j^t, \tilde{u}_j^t)$.
       %\begin{figure}[H]\centering
       \begin{figure}[htbp]\centering
        % \begin{tikzpicture}
        %  \draw[->, >=stealth, very thick] (-1.5, 0)--(1.5, 0);
        %  \draw[->, >=stealth, very thick] (0, 0.7)--(0, -0.7);
        %  \draw (-1.5, 0)node[left]{Carrier: $\tilde{u}_j^t$};
        %  \draw (1.5, 0)node[right]{$\tilde{u}_{j+1}^t$};
        %  \draw (0, 0.7)node[above]{$\underbrace{0\cdots\cdots0}_{\scalebox{1}{$\tilde{L}-\tilde{\eta}_j^t$}}\,\,\,\underbrace{1\cdots1}_{\scalebox{1}{$\tilde{\eta}_j^t$}}$};
        %  \draw (0, -0.7)node[below]{$\underbrace{1\cdots\cdots1}_{\scalebox{1}{$\tilde{\eta}_j^{t+1}=\tilde{L}-\tilde{\eta}_j^t$}}\underbrace{0\cdots0}_{\scalebox{1}{$\tilde{\eta}_j^t$}}$};
        % \end{tikzpicture}
        \includegraphics[height=3cm]{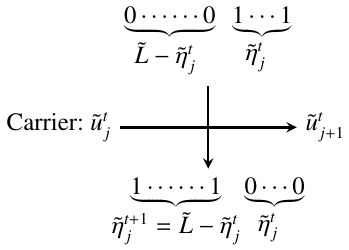}
        \caption{\label{fig2}Case 2 in Theorem \ref{th2.2}.}
       \end{figure}
      \item Case 3: $j\in I$ (in this case $\tilde{\eta}_j^t+\tilde{u}_j^t\geq \tilde{L}$)

       The carrier picks up $\tilde{\eta}_j^t$ balls. Then, the number of balls in the carrier is $\tilde{u}_j^t+\tilde{\eta}_j^t$. Since $\tilde{u}_j^t+\tilde{\eta}_j^t\geq \tilde{u}_j^t\geq \tilde{L}-\tilde{\eta}_j^t$, the carrier drops off $\tilde{L}-\tilde{\eta}_j^t$ balls as shown in Fig. \ref{fig3}. The number of balls in the boxes in this interval becomes $\tilde{L}-\tilde{\eta}_j^t$, which is equal to $\tilde{\eta}_j^{t+1}=\min(\tilde{L}-\tilde{\eta}_j^t, \tilde{u}_j^t)$.
       %\begin{figure}[H]\centering
       \begin{figure}[htbp]\centering
        % \begin{tikzpicture}
        %  \draw[->, >=stealth, very thick] (-1.5, 0)--(1.5, 0);
        %  \draw[->, >=stealth, very thick] (0, 0.7)--(0, -0.7);
        %  \draw (-1.5, 0)node[left]{Carrier: $\tilde{u}_j^t$};
        %  \draw (1.5, 0)node[right]{$\tilde{u}_{j+1}^t$};
        %  \draw (0, 0.7)node[above]{$\underbrace{1\cdots1}_{\scalebox{1}{$\,\tilde{\eta}_j^t$}}\,\,\,\,\underbrace{0\cdots\cdots0}_{\scalebox{1}{$\tilde{L}-\tilde{\eta}_j^t$}}$};
        %  \draw (0, -0.7)node[below]{$\underbrace{0\cdots0}_{\scalebox{1}{$\tilde{\eta}_j^t$}}\underbrace{1\cdots\cdots1}_{\scalebox{1}{$\tilde{L}-\tilde{\eta}_j^t=\tilde{\eta}_j^{t+1}$}}$};
        % \end{tikzpicture}
        \includegraphics[height=3cm]{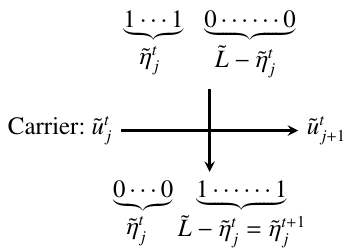}
        \caption{\label{fig3}Case 3 in Theorem \ref{th2.2}.}
       \end{figure}
     \end{itemize}
    \end{proof}
   \end{theorem}

  \subsection{Decomposition}\label{sec2.2}
Let us decompose the BBS with box capacity $L$ to the ``trivial'' dynamics, which simply moves at speed 1, and the usual Takahashi-Satsuma BBS.

   We define the following sets of semi-infinite sequences: 
   \begin{align}
    \mathcal{S}_{L, M}^{(b)} &= \left\{\BS{c}\in\mathcal{S}_{L, M}\relmiddle|c_j+c_{j+1}\leq L\,\text{for}\,j=0,1,2,\ldots\right\},\label{eq2.19}\\
    \mathcal{S}_L^{(f)} &= \{\BS{d}\in\mathcal{S}_1\mid \text{the length of every maximal subsequence that contains only 0s/1s is larger than }L\}.\label{eq2.20}
   \end{align}
   
   First, we decompose $\BS\eta\in \mathcal{S}_{L, M}$ into two sequences $\BS{c}\in\mathcal{S}_{L,M}^{(b)}$ and $\BS{d}\in\mathcal{S}_L^{(f)}$ through the decomposition map $\beta_L: \mathcal{S}_{L,M}\to\mathcal{S}_{L,M}^{(b)}\oplus\mathcal{S}_L^{(f)};\BS\eta\mapsto(\BS{c}, \BS{d})$ defined by the following procedure: 
   \begin{enumerate}
    \item Let $\tilde{\BS\eta}=(\tilde{\eta}_0, \tilde{\eta}_1, \ldots)$ be as defined above, $\tilde{\eta}_j = \eta_j + M$.
    \item Define a soliton flag sequence $\BS{a}=(a_0, a_1, \ldots)\in\mathcal{S}_1$ as 
     \begin{align}
      a_0 &= 0,\nonumber\\
      a_{j+1} &= \begin{cases}
        0 & \tilde{\eta}_j+\tilde{\eta}_{j+1} < \tilde{L},\\
       a_{j} & \tilde{\eta}_j+\tilde{\eta}_{j+1} = \tilde{L},\\
       1 & \tilde{\eta}_j+\tilde{\eta}_{j+1} > \tilde{L},
       \end{cases}\nonumber\\
       &=\begin{cases}
        0 & \eta_j+\eta_{j+1} < L,\\
       a_{j} & \eta_j+\eta_{j+1} = L,\\
       1 & \eta_j+\eta_{j+1} > L.
       \end{cases} (j=0,1,2,\ldots)\label{eq2.21}
     \end{align}
     We say $j$ is in a {\it 0-segment} if $a_j=0$, and in a {\it 1-segment} if $a_j=1$.
    \item Let $I=\{j\in\mathbb{Z}_{\geq0}\mid a_j=1, a_{j+1}=0\}$, which is the set of the right ends of consecutive {\it1-segments}. We define
    \begin{align}
     \widetilde{\rbin}(\BS\eta) = \BS{r}_I(\BS{\eta}),\label{eq2.22}
    \end{align}
    where $\BS{r}_I$ is defined in Eqs. (\ref{eq2.13}) and (\ref{eq2.14}). In Example \ref{ex2.4} below, we underline each group of $\tilde{L}$ numbers to make it easy to see the boxes.
    \item We write ``$\times$'' on the subsequence corresponding to the $j$-th and the $(j-1)$-th boxes where $j\in I$. This means that we mark the right end of every consecutive {\it 1-segment} as well as the position immediately to the left.
    \item We iteratively draw 10-arc lines from 1s in the $j$-th box without ``$\times$'' and 0s in the $(j+1)$-th box. Then, the number of arc lines that connect the $j$-th box and the $(j+1)$-th box is $\min(\tilde{\eta}_j, \tilde{L}-\tilde{\eta}_{j+1})$.
    \item Let $\tilde{\BS{b}}=(\tilde{b}_0, \tilde{b}_1, \ldots)$, where $\tilde{b}_j$ is the number of 10-arc lines that connect 1s in the $j$-th box and 0s in the $(j+1)$-th box. We will use this sequence in the proof of Theorem \ref{th2.12}. If $a_{j+1}=0$, $\tilde{\eta}_j+\tilde{\eta}_{j+1}\leq \tilde{L}$, then $\min(\tilde{\eta}_j, \tilde{L}-\tilde{\eta}_{j+1})=\tilde{\eta}_j$. If $a_{j+1}=1$, $\tilde{\eta}_j+\tilde{\eta}_{j+1}\geq \tilde{L}$, then $\min(\tilde{\eta}_j, \tilde{L}-\tilde{\eta}_{j+1})=\tilde{L}-\tilde{\eta}_{j+1}$. Thus, we get 
    \begin{align}
     \tilde{b}_j = \left\{\begin{array}{ll}
       \tilde{\eta}_j & a_j=a_{j+1}=0,\\
       \tilde{L}-\tilde{\eta}_{j+1} & a_{j+1}=a_{j+2}=1,\\
       0 & (a_j, a_{j+1})=(1, 0)\text{ or } (a_{j+1}, a_{j+2})=(1,0).
      \end{array}\right.\label{eq2.23}
    \end{align}
    \item Define a raised background sequence $\tilde{\BS{c}}=(\tilde{c}_0, \tilde{c}_1, \ldots)$ as a sequence obtained by skipping terms in the sequence $\tilde{\BS{b}}$ if ``$\times$'' is written.
    \item Define a raised soliton sequence $\tilde{\BS{d}}=(\tilde{d}_0, \tilde{d}_1, \ldots)$ as a binary sequence that is obtained by eliminating 1s and 0s connected with the 10-arc lines from $\widetilde{\rbin}(\BS{\eta})$. Let $\tilde{s}_j$ be the number of consecutive 0s and $\tilde{t}_j$ be the number of consecutive 1s in $\tilde{\BS{d}}$ as $\tilde{\BS{d}}=0^{\tilde{s}_1}1^{\tilde{t}_1}0^{\tilde{s}_2}1^{\tilde{t}_2}\cdots$.
    \item Let $\BS{c}=(c_0, c_1, \ldots)\in\mathcal{S}_{L, M}^{(b)}$, where $c_j=\tilde{c}_j-M\,(j=0, 1, \ldots)$. This sequence is called the background sequence\cite{GNN}.
    \item Let $\BS{d}=(d_0, d_1, \ldots)=0^{s_1}1^{t_1}0^{s_2}1^{t_2}\cdots$, where $s_1=\tilde{s}_1-M, s_j=\tilde{s}_j-2M\,(j=2, 3, \ldots), t_j=\tilde{t}_j-2M\,(j=1, 2, \ldots)$. We call this a soliton sequence.
   \end{enumerate}

   \begin{example}\label{ex2.4}
    $L=2$, $M=1$ and $\tilde{L}=4$.
    \begin{figure}[H]
     \centering
     \scalebox{0.86}{{\fontsize{10truept}{10truept}\selectfont\begin{tikzpicture}
      \draw node at (-0.6, 3.55)[above]{$\BS\eta$};
      \draw node at (0, 3.55)[above]{$=$};
      \draw node at (0.6, 3.55)[above]{$0$};
      \draw node at (1.4, 3.55)[above]{$0$};
      \draw node at (2.2, 3.55)[above]{$3$};
      \draw node at (3.0, 3.55)[above]{$1$};
      \draw node at (3.8, 3.55)[above]{$-1$};
      \draw node at (4.6, 3.55)[above]{$2$};
      \draw node at (5.4, 3.55)[above]{$1$};
      \draw node at (6.2, 3.55)[above]{$1$};
      \draw node at (7.0, 3.55)[above]{$0$};
      \draw node at (7.8, 3.55)[above]{$0$};
      \draw node at (8.2, 3.55)[above]{$\cdots$};
      \draw node at (-0.6, 3.05)[above]{$\BS{a}$};
      \draw node at (0, 3.05)[above]{$=$};
      \draw node at (0.6, 3.05)[above]{$0$};
      \draw node at (1.4, 3.05)[above]{$0$};
      \draw node at (2.2, 3.05)[above]{$1$};
      \draw node at (3.0, 3.05)[above]{$1$};
      \draw node at (3.8, 3.05)[above]{$0$};
      \draw node at (4.6, 3.05)[above]{$0$};
      \draw node at (5.4, 3.05)[above]{$1$};
      \draw node at (6.2, 3.05)[above]{$1$};
      \draw node at (7.0, 3.05)[above]{$0$};
      \draw node at (7.8, 3.05)[above]{$0$};
      \draw node at (8.2, 3.05)[above]{$\cdots$};
      \draw node at (-0.6, 2.55)[above]{$\tilde{\BS\eta}$};
      \draw node at (0, 2.55)[above]{$=$};
      \draw node at (0.6, 2.55)[above]{$1$};
      \draw node at (1.4, 2.55)[above]{$1$};
      \draw node at (2.2, 2.55)[above]{$4$};
      \draw node at (3.0, 2.55)[above]{$2$};
      \draw node at (3.8, 2.55)[above]{$0$};
      \draw node at (4.6, 2.55)[above]{$3$};
      \draw node at (5.4, 2.55)[above]{$2$};
      \draw node at (6.2, 2.55)[above]{$2$};
      \draw node at (7.0, 2.55)[above]{$1$};
      \draw node at (7.8, 2.55)[above]{$1$};
      \draw node at (8.2, 2.55)[above]{$\cdots$};
      \draw node at (-0.6, 1.8)[above]{$\widetilde{\rbin}(\BS\eta)$};
      \draw node at (0, 1.9)[above]{$=$};
      \draw[thick] (0.23, 1.95) -- (0.97, 1.95);
      \draw node at (0.3, 1.9)[above]{$0$};
      \draw node at (0.5, 1.9)[above]{$0$};
      \draw node at (0.7, 1.9)[above]{$0$};
      \draw node at (0.9, 1.9)[above]{$1$};
      \draw (0.9, 1.9) arc(180:360:0.1);
      \draw[thick] (1.03, 1.95) -- (1.77, 1.95);
      \draw node at (1.1, 1.9)[above]{$0$};
      \draw node at (1.3, 1.9)[above]{$0$};
      \draw node at (1.5, 1.9)[above]{$0$};
      \draw node at (1.7, 1.9)[above]{$1$};
      \draw[thick] (1.83, 1.95) -- (2.57, 1.95);
      \draw node at (1.9, 1.9)[above]{$1$};
      \draw node at (2.1, 1.9)[above]{$1$};
      \draw node at (2.3, 1.9)[above]{$1$};
      \draw node at (2.5, 1.9)[above]{$1$};
      \draw node at (2.2, 2.2)[above]{$\times$};
      \draw[thick] (2.63, 1.95) -- (3.37, 1.95);
      \draw node at (2.7, 1.9)[above]{$1$};
      \draw node at (2.9, 1.9)[above]{$1$};
      \draw node at (3.1, 1.9)[above]{$0$};
      \draw node at (3.3, 1.9)[above]{$0$};
      \draw node at (3.0, 2.2)[above]{$\times$};
      \draw[thick] (3.43, 1.95) -- (4.17, 1.95);
      \draw node at (3.5, 1.9)[above]{$0$};
      \draw node at (3.7, 1.9)[above]{$0$};
      \draw node at (3.9, 1.9)[above]{$0$};
      \draw node at (4.1, 1.9)[above]{$0$};
      \draw[thick] (4.23, 1.95) -- (4.97, 1.95);
      \draw node at (4.3, 1.9)[above]{$0$};
      \draw node at (4.5, 1.9)[above]{$1$};
      \draw node at (4.7, 1.9)[above]{$1$};
      \draw node at (4.9, 1.9)[above]{$1$};
      \draw (4.9, 1.9) arc(180:360:0.1);
      \draw (4.7, 1.9) arc(180:360:0.3);
      \draw[thick] (5.03, 1.95) -- (5.77, 1.95);
      \draw node at (5.1, 1.9)[above]{$0$};
      \draw node at (5.3, 1.9)[above]{$0$};
      \draw node at (5.5, 1.9)[above]{$1$};
      \draw node at (5.7, 1.9)[above]{$1$};
      \draw node at (5.4, 2.2)[above]{$\times$};
      \draw[thick] (5.83, 1.95) -- (6.57, 1.95);
      \draw node at (5.9, 1.9)[above]{$1$};
      \draw node at (6.1, 1.9)[above]{$1$};
      \draw node at (6.3, 1.9)[above]{$0$};
      \draw node at (6.5, 1.9)[above]{$0$};
      \draw node at (6.2, 2.2)[above]{$\times$};
      \draw[thick] (6.63, 1.95) -- (7.37, 1.95);
      \draw node at (6.7, 1.9)[above]{$0$};
      \draw node at (6.9, 1.9)[above]{$0$};
      \draw node at (7.1, 1.9)[above]{$0$};
      \draw node at (7.3, 1.9)[above]{$1$};
      \draw (7.3, 1.9) arc(180:360:0.1);
      \draw[thick] (7.43, 1.95) -- (8.17, 1.95);
      \draw node at (7.5, 1.9)[above]{$0$};
      \draw node at (7.7, 1.9)[above]{$0$};
      \draw node at (7.9, 1.9)[above]{$0$};
      \draw node at (8.1, 1.9)[above]{$1$};
      \draw (8.1, 1.9) arc(180:270:0.1);
      \draw node at (8.5, 1.9)[above]{$\cdots$};
      \draw node at (-0.6, 1.0)[above]{$\tilde{\BS{b}}$};
      \draw node at (0, 1.0)[above]{$=$};
      \draw node at (0.6, 1.0)[above]{$1$};
      \draw node at (1.4, 1.0)[above]{$0$};
      \draw node at (2.2, 1.0)[above]{$0$};
      \draw node at (3.0, 1.0)[above]{$0$};
      \draw node at (3.8, 1.0)[above]{$0$};
      \draw node at (4.6, 1.0)[above]{$2$};
      \draw node at (5.4, 1.0)[above]{$0$};
      \draw node at (6.2, 1.0)[above]{$0$};
      \draw node at (7.0, 1.0)[above]{$1$};
      \draw node at (7.8, 1.0)[above]{$1$};
      \draw node at (8.2, 1.0)[above]{$\cdots$};
      \draw node at (-0.6, 0.5)[above]{$\tilde{\BS{c}}$};
      \draw node at (0, 0.5)[above]{$=$};
      \draw node at (0.6, 0.5)[above]{$1$};
      \draw node at (1.4, 0.5)[above]{$0$};
      \draw node at (3.8, 0.5)[above]{$0$};
      \draw node at (4.6, 0.5)[above]{$2$};
      \draw node at (7.0, 0.5)[above]{$1$};
      \draw node at (7.8, 0.5)[above]{$1$};
      \draw node at (8.2, 0.5)[above]{$\cdots$};
      \draw node at (-0.6, 0)[above]{$\BS{c}$};
      \draw node at (0, 0)[above]{$=$};
      \draw node at (0.6, 0)[above]{$0$};
      \draw node at (1.4, 0)[above]{$-1$};
      \draw node at (3.8, 0)[above]{$-1$};
      \draw node at (4.6, 0)[above]{$1$};
      \draw node at (7.0, 0)[above]{$0$};
      \draw node at (7.8, 0)[above]{$0$};
      \draw node at (8.2, 0)[above]{$\cdots$};
      \draw node at (-0.6, -0.5)[above]{$\tilde{\BS{d}}$};
      \draw node at (0, -0.5)[above]{$=$};
      \draw node at (0.4, -0.5)[above]{$0$};
      \draw node at (0.6, -0.5)[above]{$0$};
      \draw node at (0.8, -0.5)[above]{$0$};
      \draw node at (1.2, -0.5)[above]{$0$};
      \draw node at (1.4, -0.5)[above]{$0$};
      \draw node at (1.6, -0.5)[above]{$1$};
      \draw node at (1.9, -0.5)[above]{$1$};
      \draw node at (2.1, -0.5)[above]{$1$};
      \draw node at (2.3, -0.5)[above]{$1$};
      \draw node at (2.5, -0.5)[above]{$1$};
      \draw node at (2.7, -0.5)[above]{$1$};
      \draw node at (2.9, -0.5)[above]{$1$};
      \draw node at (3.1, -0.5)[above]{$0$};
      \draw node at (3.3, -0.5)[above]{$0$};
      \draw node at (3.5, -0.5)[above]{$0$};
      \draw node at (3.7, -0.5)[above]{$0$};
      \draw node at (3.9, -0.5)[above]{$0$};
      \draw node at (4.1, -0.5)[above]{$0$};
      \draw node at (4.5, -0.5)[above]{$0$};
      \draw node at (4.7, -0.5)[above]{$1$};
      \draw node at (5.3, -0.5)[above]{$1$};
      \draw node at (5.5, -0.5)[above]{$1$};
      \draw node at (5.9, -0.5)[above]{$1$};
      \draw node at (6.1, -0.5)[above]{$1$};
      \draw node at (6.3, -0.5)[above]{$0$};
      \draw node at (6.5, -0.5)[above]{$0$};
      \draw node at (6.8, -0.5)[above]{$0$};
      \draw node at (7.0, -0.5)[above]{$0$};
      \draw node at (7.2, -0.5)[above]{$0$};
      \draw node at (7.7, -0.5)[above]{$0$};
      \draw node at (7.9, -0.5)[above]{$0$};
      \draw node at (8.3, -0.5)[above]{$\cdots$};
      \draw node at (0, -1.0)[above]{$=$};
      \draw node at (0.3, -1.0)[above]{$0^5$};
      \draw node at (0.65, -1.0)[above]{$1^7$};
      \draw node at (1.0, -1.0)[above]{$0^7$};
      \draw node at (1.35, -1.0)[above]{$1^5$};
      \draw node at (1.75, -1.0)[above]{$0^\infty$};
      \draw node at (-0.6, -1.5)[above]{$\BS{d}$};
      \draw node at (0, -1.5)[above]{$=$};
      \draw node at (0.3, -1.5)[above]{$0^4$};
      \draw node at (0.65, -1.5)[above]{$1^5$};
      \draw node at (1.0, -1.5)[above]{$0^5$};
      \draw node at (1.35, -1.5)[above]{$1^3$};
      \draw node at (1.75, -1.5)[above]{$0^\infty$};
     \end{tikzpicture}}}
    \end{figure}
   \end{example}
  
   Define the forward-shift operator $\Lambda$ on $\mathcal{S}_{L, M}^{(b)}$ as
   \begin{align}
    (\Lambda(\BS{c}))_j = \left\{\begin{array}{ll}
     0 & j=0,\\
     (\BS{c})_{j-1} & \text{otherwise}.
    \end{array}\right.\label{eq2.24}
   \end{align}
  
   \begin{theorem}\label{th2.5}
    For $\BS\eta\in\mathcal{S}_{L, M}$, let $\BS{c}(\BS\eta)$ be the background sequence and $\BS{d}(\BS\eta)$ be the soliton sequence, as defined above.
    \begin{align}
     \Lambda(\BS{c}(\BS{\eta})) &= \BS{c}(T_L(\BS{\eta}))\label{eq2.25}\\
     T_1(\BS{d}(\BS{\eta})) &= \BS{d}(T_L(\BS{\eta}))\label{eq2.26}
    \end{align}
   \end{theorem}
  
   The background sequence $\BS{c}$ is shifted to the right at speed 1, and the soliton sequence $\BS{d}$ follows the time evolution of BBS($1$). Because the time evolution of BBS($1$) can be linearized \cite{KNTW}, we can linearize the time evolution of BBS($L$). This claim will be proved from Theorems \ref{th2.9} and \ref{th2.12}. 
  
   \begin{example}\label{ex2.6}
    $L=2$, and $\BS\eta=0,0,3,1,-1,2,1,1,0,0,\ldots\in\mathcal{S}_{2, 1}$ (as in Example \ref{ex2.4})

    \begin{align*}
     \BS{c}(\BS\eta) &= 0, -1, -1, 1, 0, 0, \ldots\\
     \BS{d}(\BS\eta) &= 0000111110000011100000\cdots\\
     T_2(\BS\eta) &= 0, 0, -1, 1, 3, 0, 1, 1, 2, 0, \ldots\\
     \BS{c}(T_2(\BS\eta)) &= 0, 0, -1, -1, 1, 0, \ldots\\
     \BS{d}(T_2(\BS\eta)) &= 0000000001111100011100\cdots
    \end{align*}
    \begin{figure}[H]
     \centering
     \begin{tikzpicture}[auto, ->]
      \node (a) at (0, 2) {$\BS\eta$};
      \node (x) at (7, 2) {$T_2(\BS\eta)$};
      \node (b) at (0, 0) {$\left\{\begin{array}{l}\BS{c}(\BS\eta)\\\BS{d}(\BS\eta)\end{array}\right\}$};
      \node (y) at (7, 0) {$\left\{\begin{array}{ll}\BS{c}(T_2(\BS\eta))\\\BS{d}(T_2(\BS\eta))\end{array}\right\}$};
      \draw (a) -- node {$T_2$} (x);
      \draw (x) -- node {$\beta_2$} (y);
      \draw (a) -- node[swap] {$\beta_2$} (b);
      \draw (b) -- node[swap, align=left] {$\BS{c}(T_2(\BS\eta))=\Lambda(\BS{c}(\BS\eta))$\\ $\BS{d}(T_2(\BS\eta))=T_1(\BS{d}(\BS\eta))$} (y);
     \end{tikzpicture}
    \end{figure}
   \end{example}
     
   To prove Theorem \ref{th2.5}, we define another procedure to obtain the background sequence and the soliton sequence.

   \begin{enumerate}
    \item Define a soliton flag sequence $\BS{a}^\prime=(a_{-1}^\prime, a_0^\prime, \ldots)\in\mathcal{S}_1$ as 
     \begin{align}
      a_{-1}^\prime &= 0,\nonumber\\
      a_{j+1}^\prime &= \begin{cases}
        0 & \tilde{\eta}_{j+1}+\tilde{\eta}_{j+2} < \tilde{L},\\
       a_{j}^\prime & \tilde{\eta}_{j+1}+\tilde{\eta}_{j+2} = \tilde{L},\\
       1 & \tilde{\eta}_{j+1}+\tilde{\eta}_{j+2} > \tilde{L},
       \end{cases}\nonumber\\
       &=\begin{cases}
        0 & \eta_{j+1}+\eta_{j+2} < L,\\
       a_{j} & \eta_{j+1}+\eta_{j+2} = L,\\
       1 & \eta_{j+1}+\eta_{j+2} > L.
       \end{cases} (j=0,1,2,\ldots)\label{eq2.27}
     \end{align}
     We say $j$ is in a {\it right-0-segment} if $a_j^\prime=0$, and in a {\it right-1-segment} if $a_j^\prime=1$.
     By definition, it is clear that
     \begin{align}
      a_j^\prime &= a_{j+1}\quad(j=0, 1, \ldots).\label{eq2.28}
     \end{align}
    \item Let $I^\prime=\{j\in\mathbb{Z}_{\geq0}\mid a_j^\prime=1, a_{j-1}^\prime=0\}$, which is the set of the left ends of consecutive {\it right-1-segments}. We define
    \begin{align}
     \widetilde{\lbin}(\BS\eta) = \BS{l}_{I^\prime}(\tilde{\BS\eta}),\label{eq2.29}
    \end{align}
    where $\BS{l}_{I^\prime}$ is defined in Eqs. (\ref{eq2.16}) and (\ref{eq2.17}). 
    \item We write ``$\times$'' on the subsequence corresponding to the $j$-th and the $(j+1)$-th boxes where $j\in I$. This means that we mark the left end of every consecutive {\it right-1-segment} as well as the position immediately to the right.
    \item We iteratively draw 01-arc lines from 0s in the $(j-1)$-th box without ``$\times$'' and 1s in the $j$-th box repeatedly. Then, the number of arc lines that connect the $(j-1)$-th box and the $j$-th box is $\min(\eta_j, L-\eta_{j-1})$.
    \item Let $\tilde{\BS{b}}^\prime=(\tilde{b}_0^\prime, \tilde{b}_1^\prime, \ldots)$, where $\tilde{b}_j^\prime$ is the number of 01-arc lines that connect 0s in the $(j-1)$-th box and 1s in the $j$-th box. If $a_{j-1}^\prime=0$, $\tilde{\eta}_{j-1}+\tilde{\eta}_{j}\leq \tilde{L}$, then $\min(\tilde{\eta}_j, \tilde{L}-\tilde{\eta}_{j-1})=\tilde{\eta}_j$. If $a_{j-1}=1$, $\tilde{\eta}_{j-1}+\tilde{\eta}_{j}\geq \tilde{L}$, then $\min(\tilde{\eta}_j, \tilde{L}-\tilde{\eta}_{j-1})=\tilde{L}-\tilde{\eta}_{j-1}$. Thus, we get 
    \begin{align}
     \tilde{b}_j^\prime = \begin{cases}
       \tilde{\eta}_j & a_{j-1}^\prime=a_{j}^\prime=0,\\
       \tilde{L}-\tilde{\eta}_{j-1} & a_{j-2}^\prime=a_{j-1}^\prime=1,\\
       0 & (a_{j-2}^\prime, a_{j-1}^\prime)=(0, 1)\text{ or } (a_{j-1}^\prime, a_{j}^\prime)=(0, 1).
      \end{cases}\label{eq2.30}
    \end{align}
    \item Let $\tilde{\BS{c}}^\prime=(\tilde{c}_0^\prime, \tilde{c}_1^\prime, \ldots)$ be a sequence obtained by skipping terms in the sequence $\tilde{\BS{b}}^\prime$ if ``$\times$'' is written.
    \item Let $\tilde{\BS{d}}^\prime=(\tilde{d}_0^\prime, \tilde{d}_1^\prime, \ldots)$ be a binary sequence that is obtained by eliminating 1s and 0s connected with the 01-arc lines from $\widetilde{\lbin}(\BS{\eta})$. Let $\tilde{s}_j^\prime$ be the number of consecutive 0s and $\tilde{t}_j^\prime$ be the number of consecutive 1s in $\tilde{\BS{d}}^\prime$ as $\tilde{\BS{d}}^\prime=0^{\tilde{s}_1^\prime}1^{\tilde{t}_1^\prime}0^{\tilde{s}_2^\prime}1^{\tilde{t}_2^\prime}\cdots$.

    \item Let $\BS{c}^\prime=(c_0^\prime, c_1^\prime, \ldots)\in\mathcal{S}_{L, M}^{(b)}$, where $c_j^\prime=\tilde{c}_j^\prime-M\,(j=0, 1, \ldots)$.
    \item Let $\BS{d}^\prime=(d_0^\prime, d_1^\prime, \ldots)=0^{s_1^\prime}1^{t_1^\prime}0^{s_2^\prime}1^{t_2^\prime}\cdots$, where $s_j^\prime=\tilde{s}_j^\prime-2M, t_j^\prime=\tilde{t}_j^\prime-2M\,(j=1, 2, \ldots)$.
   \end{enumerate}
   \begin{example}\label{ex2.11}
    $L=2$, $\BS\eta=0, 0, 3, 1, -1, 2, 1, 1, 0, 0, \ldots\in\mathcal{S}_{2, 1}$ (as in Examples \ref{ex2.4} and \ref{ex2.6}.)
    \begin{figure}[H]
     \centering
     \scalebox{0.86}{{\fontsize{10truept}{10truept}\selectfont\begin{tikzpicture}
      \draw node at (-0.65, 3.55)[above]{$T_2(\BS\eta)$};
      \draw node at (0, 3.55)[above]{$=$};
      \draw node at (0.6, 3.55)[above]{$0$};
      \draw node at (1.4, 3.55)[above]{$0$};
      \draw node at (2.2, 3.55)[above]{$-1$};
      \draw node at (3.0, 3.55)[above]{$1$};
      \draw node at (3.8, 3.55)[above]{$3$};
      \draw node at (4.6, 3.55)[above]{$0$};
      \draw node at (5.4, 3.55)[above]{$1$};
      \draw node at (6.2, 3.55)[above]{$1$};
      \draw node at (7.0, 3.55)[above]{$2$};
      \draw node at (7.8, 3.55)[above]{$0$};
      \draw node at (8.2, 3.55)[above]{$\cdots$};
      \draw node at (-0.8, 3.05)[above]{$\BS{a}^\prime(T_2(\BS\eta))$};
      \draw node at (0, 3.05)[above]{$=$};
      \draw node at (0.6, 3.05)[above]{$0$};
      \draw node at (1.4, 3.05)[above]{$0$};
      \draw node at (2.2, 3.05)[above]{$0$};
      \draw node at (3.0, 3.05)[above]{$1$};
      \draw node at (3.8, 3.05)[above]{$1$};
      \draw node at (4.6, 3.05)[above]{$0$};
      \draw node at (5.4, 3.05)[above]{$0$};
      \draw node at (6.2, 3.05)[above]{$1$};
      \draw node at (7.0, 3.05)[above]{$1$};
      \draw node at (7.8, 3.05)[above]{$0$};
      \draw node at (8.2, 3.05)[above]{$\cdots$};
      \draw node at (-0.6, 2.55)[above]{$\tilde{T_2(\BS\eta)}$};
      \draw node at (0, 2.55)[above]{$=$};
      \draw node at (0.6, 2.55)[above]{$1$};
      \draw node at (1.4, 2.55)[above]{$1$};
      \draw node at (2.2, 2.55)[above]{$0$};
      \draw node at (3.0, 2.55)[above]{$2$};
      \draw node at (3.8, 2.55)[above]{$4$};
      \draw node at (4.6, 2.55)[above]{$1$};
      \draw node at (5.4, 2.55)[above]{$2$};
      \draw node at (6.2, 2.55)[above]{$2$};
      \draw node at (7.0, 2.55)[above]{$3$};
      \draw node at (7.8, 2.55)[above]{$1$};
      \draw node at (8.2, 2.55)[above]{$\cdots$};
      \draw node at (-0.87, 1.8)[above]{$\widetilde{\lbin}(T_2(\BS\eta))$};
      \draw node at (0, 1.9)[above]{$=$};
      \draw[thick] (0.23, 1.95) -- (0.97, 1.95);
      \draw (0.3, 1.9) arc(360:270:0.1);
      \draw node at (0.3, 1.9)[above]{$1$};
      \draw node at (0.5, 1.9)[above]{$0$};
      \draw node at (0.7, 1.9)[above]{$0$};
      \draw node at (0.9, 1.9)[above]{$0$};
      \draw[thick] (1.03, 1.95) -- (1.77, 1.95);
      \draw (0.9, 1.9) arc(180:360:0.1);
      \draw node at (1.1, 1.9)[above]{$1$};
      \draw node at (1.3, 1.9)[above]{$0$};
      \draw node at (1.5, 1.9)[above]{$0$};
      \draw node at (1.7, 1.9)[above]{$0$};
      \draw[thick] (1.83, 1.95) -- (2.57, 1.95);
      \draw node at (1.9, 1.9)[above]{$0$};
      \draw node at (2.1, 1.9)[above]{$0$};
      \draw node at (2.3, 1.9)[above]{$0$};
      \draw node at (2.5, 1.9)[above]{$0$};
      \draw node at (2.2, 2.2)[above]{$\times$};
      \draw[thick] (2.63, 1.95) -- (3.37, 1.95);
      \draw node at (2.7, 1.9)[above]{$0$};
      \draw node at (2.9, 1.9)[above]{$0$};
      \draw node at (3.1, 1.9)[above]{$1$};
      \draw node at (3.3, 1.9)[above]{$1$};
      \draw node at (3.0, 2.2)[above]{$\times$};
      \draw[thick] (3.43, 1.95) -- (4.17, 1.95);
      \draw node at (3.5, 1.9)[above]{$1$};
      \draw node at (3.7, 1.9)[above]{$1$};
      \draw node at (3.9, 1.9)[above]{$1$};
      \draw node at (4.1, 1.9)[above]{$1$};
      \draw[thick] (4.23, 1.95) -- (4.97, 1.95);
      \draw node at (4.3, 1.9)[above]{$1$};
      \draw node at (4.5, 1.9)[above]{$0$};
      \draw node at (4.7, 1.9)[above]{$0$};
      \draw node at (4.9, 1.9)[above]{$0$};
      \draw (4.9, 1.9) arc(180:360:0.1);
      \draw (4.7, 1.9) arc(180:360:0.3);
      \draw[thick] (5.03, 1.95) -- (5.77, 1.95);
      \draw node at (5.1, 1.9)[above]{$1$};
      \draw node at (5.3, 1.9)[above]{$1$};
      \draw node at (5.5, 1.9)[above]{$0$};
      \draw node at (5.7, 1.9)[above]{$0$};
      \draw node at (5.4, 2.2)[above]{$\times$};
      \draw[thick] (5.83, 1.95) -- (6.57, 1.95);
      \draw node at (5.9, 1.9)[above]{$0$};
      \draw node at (6.1, 1.9)[above]{$0$};
      \draw node at (6.3, 1.9)[above]{$1$};
      \draw node at (6.5, 1.9)[above]{$1$};
      \draw node at (6.2, 2.2)[above]{$\times$};
      \draw[thick] (6.63, 1.95) -- (7.37, 1.95);
      \draw node at (6.7, 1.9)[above]{$1$};
      \draw node at (6.9, 1.9)[above]{$1$};
      \draw node at (7.1, 1.9)[above]{$1$};
      \draw node at (7.3, 1.9)[above]{$0$};
      \draw (7.3, 1.9) arc(180:360:0.1);
      \draw[thick] (7.43, 1.95) -- (8.17, 1.95);
      \draw node at (7.5, 1.9)[above]{$1$};
      \draw node at (7.7, 1.9)[above]{$0$};
      \draw node at (7.9, 1.9)[above]{$0$};
      \draw node at (8.1, 1.9)[above]{$0$};
      \draw (8.1, 1.9) arc(180:270:0.1);
      \draw node at (8.5, 1.9)[above]{$\cdots$};
      \draw node at (-0.8, 1.0)[above]{$\tilde{\BS{b}}^\prime(T_2(\BS\eta))$};
      \draw node at (0, 1.0)[above]{$=$};
      \draw node at (0.6, 1.0)[above]{$1$};
      \draw node at (1.4, 1.0)[above]{$1$};
      \draw node at (2.2, 1.0)[above]{$0$};
      \draw node at (3.0, 1.0)[above]{$0$};
      \draw node at (3.8, 1.0)[above]{$0$};
      \draw node at (4.6, 1.0)[above]{$0$};
      \draw node at (5.4, 1.0)[above]{$2$};
      \draw node at (6.2, 1.0)[above]{$0$};
      \draw node at (7.0, 1.0)[above]{$0$};
      \draw node at (7.8, 1.0)[above]{$1$};
      \draw node at (8.2, 1.0)[above]{$\cdots$};
      \draw node at (-0.78, 0.5)[above]{$\tilde{\BS{c}}^\prime(T_2(\BS\eta))$};
      \draw node at (0, 0.5)[above]{$=$};
      \draw node at (0.6, 0.5)[above]{$1$};
      \draw node at (1.4, 0.5)[above]{$1$};
      \draw node at (2.2, 0.5)[above]{$0$};
      \draw node at (4.6, 0.5)[above]{$0$};
      \draw node at (5.4, 0.5)[above]{$2$};
      \draw node at (7.8, 0.5)[above]{$1$};
      \draw node at (8.2, 0.5)[above]{$\cdots$};
      \draw node at (-0.78, 0)[above]{$\BS{c}^\prime(T_2(\BS\eta))$};
      \draw node at (0, 0)[above]{$=$};
      \draw node at (0.6, 0)[above]{$0$};
      \draw node at (1.4, 0)[above]{$0$};
      \draw node at (2.2, 0)[above]{$-1$};
      \draw node at (4.6, 0)[above]{$-1$};
      \draw node at (5.4, 0)[above]{$1$};
      \draw node at (7.8, 0)[above]{$0$};
      \draw node at (8.2, 0)[above]{$\cdots$};
      \draw node at (-0.78, -0.5)[above]{$\tilde{\BS{d}}^\prime(T_2(\BS\eta))$};
      \draw node at (0, -0.5)[above]{$=$};
      \draw node at (0.5, -0.5)[above]{$0$};
      \draw node at (0.7, -0.5)[above]{$0$};
      \draw node at (1.2, -0.5)[above]{$0$};
      \draw node at (1.4, -0.5)[above]{$0$};
      \draw node at (1.6, -0.5)[above]{$0$};
      \draw node at (1.9, -0.5)[above]{$0$};
      \draw node at (2.1, -0.5)[above]{$0$};
      \draw node at (2.3, -0.5)[above]{$0$};
      \draw node at (2.5, -0.5)[above]{$0$};
      \draw node at (2.7, -0.5)[above]{$0$};
      \draw node at (2.9, -0.5)[above]{$0$};
      \draw node at (3.1, -0.5)[above]{$1$};
      \draw node at (3.3, -0.5)[above]{$1$};
      \draw node at (3.5, -0.5)[above]{$1$};
      \draw node at (3.7, -0.5)[above]{$1$};
      \draw node at (3.9, -0.5)[above]{$1$};
      \draw node at (4.1, -0.5)[above]{$1$};
      \draw node at (4.5, -0.5)[above]{$1$};
      \draw node at (4.7, -0.5)[above]{$0$};
      \draw node at (5.3, -0.5)[above]{$0$};
      \draw node at (5.5, -0.5)[above]{$0$};
      \draw node at (5.9, -0.5)[above]{$0$};
      \draw node at (6.1, -0.5)[above]{$0$};
      \draw node at (6.3, -0.5)[above]{$1$};
      \draw node at (6.5, -0.5)[above]{$1$};
      \draw node at (6.8, -0.5)[above]{$1$};
      \draw node at (7.0, -0.5)[above]{$1$};
      \draw node at (7.2, -0.5)[above]{$1$};
      \draw node at (7.7, -0.5)[above]{$0$};
      \draw node at (7.9, -0.5)[above]{$0$};
      \draw node at (8.3, -0.5)[above]{$\cdots$};
      \draw node at (0, -1.0)[above]{$=$};
      \draw node at (0.33, -1.0)[above]{$0^{11}$};
      \draw node at (0.65, -1.0)[above]{$1^7$};
      \draw node at (1.0, -1.0)[above]{$0^5$};
      \draw node at (1.35, -1.0)[above]{$1^5$};
      \draw node at (1.75, -1.0)[above]{$0^\infty$};
      \draw node at (-0.78, -1.5)[above]{$\BS{d}^\prime$};
      \draw node at (0, -1.5)[above]{$=$};
      \draw node at (0.3, -1.5)[above]{$0^9$};
      \draw node at (0.65, -1.5)[above]{$1^5$};
      \draw node at (1.0, -1.5)[above]{$0^3$};
      \draw node at (1.35, -1.5)[above]{$1^3$};
      \draw node at (1.75, -1.5)[above]{$0^\infty$};
     \end{tikzpicture}}}
    \end{figure}
   \end{example}

   Examples \ref{ex2.4} and \ref{ex2.11} imply the following Lemmas \ref{lem2.7}, \ref{lem2.8},  and Theorem \ref{th2.9}.
    
   \begin{lemma} \label{lem2.7}
    For $\BS{\eta}\in\mathcal{S}_{L,M}$, the following conditions are equivalent.
    \begin{enumerate}\renewcommand{\labelenumi}{(\roman{enumi})}
     \item $(\BS{a}(\BS{\eta}))_j=1$ and $(\BS{a}(\BS{\eta}))_{j+1}=0$
     \item $(\BS{a}^\prime(T_L(\BS{\eta})))_j=1$ and $(\BS{a}^\prime(T_L(\BS{\eta})))_{j-1}=0$
    \end{enumerate}
    \begin{proof}
     Let $\BS{a}(\BS{\eta})=(a_0, a_1, \ldots), \BS{a}^\prime(T_L(\BS{\eta}))=(a_0^\prime, a_1^\prime, \ldots)$ . Here, we prove that (i) $\Rightarrow$ (ii) by using Eqs. (\ref{eq2.3}) and (\ref{eq2.4}). The reverse (ii) $\Rightarrow$ (i) can be proved similarly.
     \begin{itemize}
      \item Case 1: $\eta_{j-1}^t+\eta_j^t>L, \eta_j^t+\eta_{j+1}^t<L$ (See Fig. \ref{fig4}.)

       $\eta_j^t+u_j^t-L\geq \eta_j^t+\eta_{j-1}^t-L>0$, and
       \begin{align*}
        \eta_j^{t+1} &= \min(u_j^t, L-\eta_j^t)\\
         &= L-\eta_j^t.
       \end{align*}
       \begin{figure}[H]\centering
        \includegraphics[width=6cm]{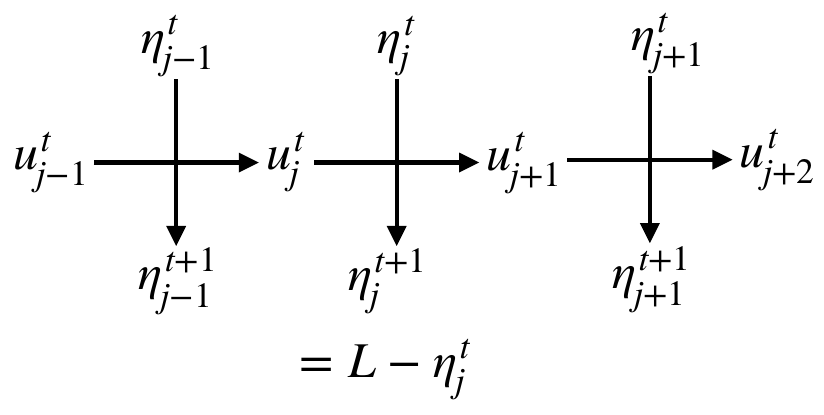}
        \caption{\label{fig4}Case 1 in Lemma \ref{lem2.7}.}
       \end{figure}
       Therefore, we have
       \begin{align*}
        \eta_{j-1}^{t+1}+\eta_j^{t+1}-L &= \min(u_{j-1}^t, L-\eta_{j-1}^t)+(L-\eta_j^t)-L\\
         &= \min(u_{j-1}^t-\eta_j^t, L-\eta_{j-1}^t-\eta_j^t)\\
         &< 0.
       \end{align*}
       By using $u_{j+1}^t=\eta_j^t+\max(0, \eta_j^t+u_j^t-L)>\eta_j^t$, 
       \begin{align*}
        \eta_j^{t+1}+\eta_{j+1}^{t+1}-L &= (L-\eta_j^t)+\min(u_{j+1}^t, L-\eta_{j+1}^t)-L\\
         &= \min(u_{j+1}^t-\eta_j^t, L-\eta_j^t-\eta_{j+1}^t)\\
         &>0.
       \end{align*}
      \item Case 2: The case $\eta_{k-1}^t+\eta_k^t>L, \eta_{i-1}^t+\eta_i^t=L\,(i=k+1, \ldots, j), \eta_j^t+\eta_{j+1}^t<L$ (See Fig. \ref{fig5}.)

       $\eta_i^t+u_i^t-L\geq \eta_i^t+\eta_{i-1}^t-L\geq0\,\text{for}\,i=k, k+1, \ldots, j$ and
       \begin{align*}
        \eta_i^{t+1} &= \min(u_j^t, L-\eta_j^t)\\
         &= L-\eta_i^t\,(i=k, k+1,  \ldots, j). 
       \end{align*}
       From $\eta_k^t+u_k^t-L\geq\eta_k^t+\eta_{k-1}^t-L>0$, we have $u_{k+1}^t>\eta_k^t$. Then, we have $\eta_{k+1}^t+u_{k+1}^t-L>\eta_{k+1}^t+\eta_k^t-L=0$, and $u_{k+2}^t>\eta_{k+1}^t$. Through an induction procedure, we also obtain $u_{j+1}^t>\eta_j^t$. 
       \begin{figure}[H]\centering
        \includegraphics[width=8.2cm]{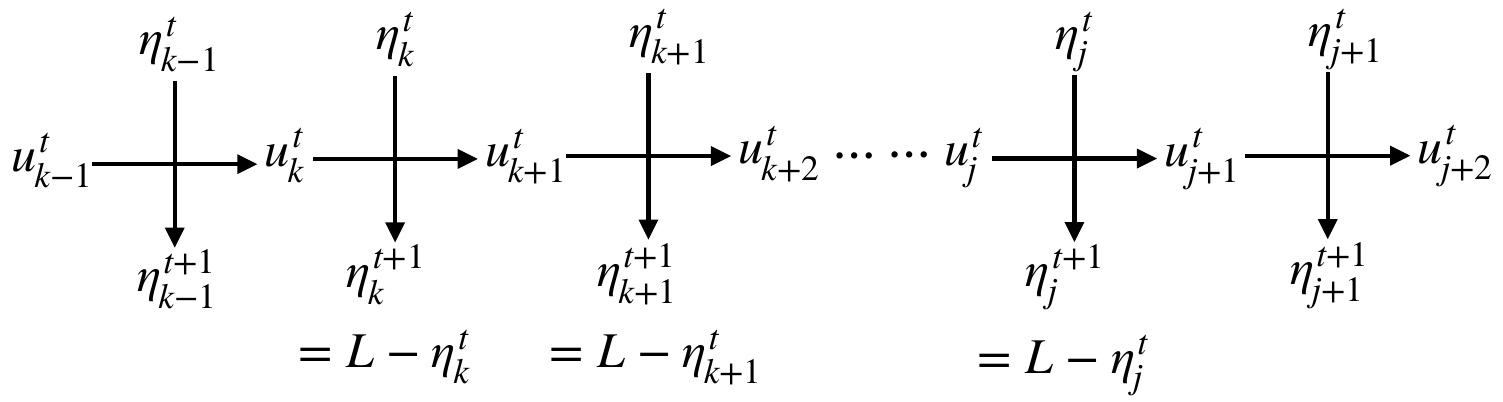}
        \caption{\label{fig5}Case 2 in Lemma \ref{lem2.7}.}
       \end{figure}
       Therefore, we have 
       \begin{align*}
        \eta_{k-1}^{t+1}+\eta_k^{t+1}-L &= \min(u_{k-1}^t, L-\eta_{k-1}^t)+(L-\eta_k^t)-L\\
         &= \min(u_{k-1}^t-\eta_k^t, L-\eta_{k-1}^t-\eta_k^t)\\
         &< 0
       \end{align*}
       \begin{align*}
        \eta_{i-1}^{t+1}+\eta_i^{t+1}-L &= (L-\eta_{i-1}^t) + (L-\eta_i^t)-L\\
         &= L-\eta_{i-1}^t-\eta_i^t\\
         &= 0\quad(i=k+1, \ldots, j)
       \end{align*}
       \begin{align*}
        \eta_j^{t+1}+\eta_{j+1}^{t+1}-L &= (L-\eta_j^t) + \min(u_{j+1}^t, L-\eta_{j+1}^t)-L\\
         &= \min(u_{j+1}^t-\eta_j^t, L-\eta_{j+1}^t-\eta_j^t)\\
         &>0.
       \end{align*}
     \end{itemize}
    \end{proof}
   \end{lemma}
    
   \begin{lemma} \label{lem2.8}
    For $\BS{\eta}\in\mathcal{S}_{L,M}$, 
    \begin{align}
     T_1^M(\widetilde{\rbin}(\BS{\eta})) &= \widetilde{\lbin}(T_L(\BS{\eta}))\label{eq2.31}
    \end{align}
    \begin{proof}
     Let $I$ be the set of indices
     \begin{align*}
      I &= \{i\in\mathbb{Z}_{\geq0}\mid (\BS{a}(\BS{\eta}))_i=1, (\BS{a}(\BS{\eta}))_{i+1}=0\}\\
       &\subset\{i\in\mathbb{Z}_{\geq0}\mid \eta_i+u_i\geq L\}.
     \end{align*}
     Then, from Lemma \ref{lem2.7},
     \begin{align*}
      I &= \{i\in\mathbb{Z}_{\geq0}\mid  (\BS{a}^\prime(T_L(\BS{\eta})))_{i-1}=0, (\BS{a}^\prime(T_L(\BS{\eta})))_i=1\}.
     \end{align*}
     From Theorem \ref{th2.2}, we have $T_1^M(\widetilde{\rbin}(\BS{\eta})) = {\widetilde{\lbin}}(T_L(\BS{\eta}))$.
    \end{proof}
   \end{lemma}
    
   \begin{theorem}\label{th2.9}
    For $\BS{\eta}\in\mathcal{S}_{L, M}$ 
    \begin{align}
     \Lambda(\BS{c}(\BS{\eta})) &= \BS{c}^\prime(T_L(\BS{\eta}))\label{eq2.32}\\
     T_1(\BS{d}(\BS{\eta})) &= \BS{d}^\prime(T_L(\BS{\eta}))\label{eq2.33}
    \end{align}
   \end{theorem}
     
   To prove Theorem \ref{th2.9}, we introduce 10-arc lines on the binary sequence $\BS{\eta}\in\mathcal{S}_1$ \cite{YYT} which can express the time evolution of BBS($1$) according to the following rules.
   \begin{enumerate}
    \renewcommand{\labelenumi}{\roman{enumi})}
    \item For $\BS\eta\in\mathcal{S}_1$, connect all 10 pairs with arc lines.
    \item Neglecting the 1s and 0s which were connected already, connect all the remaining 10 pairs with arc lines.
    \item Repeat the above procedure until all the 1s are connected to 0s.
    \item $T_1(\BS\eta)$ is the state obtained by exchanging the 1s and 0s in every connected 10 pair.
   \end{enumerate}
   We can draw 01 arc lines in the same fashion, and the following lemma \cite{KNTW} is obvious from the definition.
   \begin{lemma}\label{lem2.10}
    The 10-arc lines for $\BS\eta\in\mathcal{S}_1$ coincide with the 01-arc lines for $T_1(\BS\eta)$.
   \end{lemma}
     
   (\textit{Proof of Theorem \ref{th2.9}.})
   From Lemmas \ref{lem2.7} and \ref{lem2.8}, the 10-arc lines for $\widetilde{\rbin}(\BS\eta)$ coincide with the 01-arc lines for $\widetilde{\lbin}(T_L(\BS\eta))$. Using the 01/10-arc lines, $\min(\tilde{\eta}_j, \tilde{L}-\tilde{\eta}_{j+1})$ can be described as the number of 10-arc lines on $\widetilde{\rbin}(\BS\eta)$ that connect the balls in \textit{the $j$-th box} and the vacancies in \textit{the $(j+1)$-th box}. Similarly, $\min(\widetilde{(T_{L}(\BS{\eta}))}_j, \tilde{L}-\widetilde{(T_L(\BS\eta))}_{j-1})$ can be explained as the number of 01-arc lines on $\widetilde{\lbin}(T_L(\BS\eta))$ that connect the balls in \textit{the $j$-th box} and the vacancies in \textit{the $(j-1)$-th box}. Thus, it follows that $\Lambda(\tilde{\BS{b}}(\BS\eta)) = \tilde{\BS{b}}^\prime(T_L(\BS\eta))$ and $\Lambda(\tilde{\BS{c}}(\BS\eta)) = \tilde{\BS{c}}^\prime(T_L(\BS\eta))$.
      
   When we obtain the raised soliton sequences $\tilde{\BS{d}}(\BS\eta), \tilde{\BS{d}}^\prime(T_L(\BS\eta))$ deleting 1s, 0s and 10/01-arc lines that correspond to background materials, other 10-arc lines on $\widetilde{\rbin}(\BS\eta)$ and 01-arc lines on $\widetilde{\lbin}(T_L(\BS\eta))$ do not change. Thus, we obtain $T_1^M(\tilde{\BS{d}}(\BS\eta))=\tilde{\BS{d}}^\prime(T_L(\BS\eta))$, and $T_1(\BS{d}(\BS{\eta}))=\BS{d}^\prime(T_L(\BS\eta))$.

   \begin{theorem}\label{th2.12}
    \begin{align}
     \BS{c}(\BS{\eta}) &= \BS{c}^\prime(\BS{\eta})\label{eq2.34}\\
     \BS{d}(\BS{\eta}) &= \BS{d}^\prime(\BS{\eta})\label{eq2.35}
    \end{align}
    \begin{proof}
     For the soliton flag sequence $\BS{a}$, define indices $i_1, i_2, \ldots, i_N, k_1, k_2, \ldots, k_N\in\mathbb{Z}_{\geq0}$ as 
     \begin{align}
      \{j\in\mathbb{Z}_{\geq0}\mid a_j=0, a_{j+1}=1\} &= \{i_1<i_2<\ldots<i_N\}\label{eq2.36}\\
      \{j\in\mathbb{Z}_{\geq0}\mid a_j=1, a_{j+1}=0\} &= \{k_1<k_2<\ldots<k_N\}.\label{eq2.37}
     \end{align}
     These indices satisfy the interlacing condition $0<i_1<k_1<i_2<k_2<\cdots<i_N<k_N$. Similarly, for the soliton flag sequence $\BS{a}^\prime$, define indices $i_1^\prime, i_2^\prime, \ldots, i_N^\prime, k_1^\prime, k_2^\prime, \ldots, k_N^\prime\in\mathbb{Z}_{\geq0}$ as 
     \begin{align}
      \{j\in\mathbb{Z}_{\geq0}\mid a_{j-1}^\prime=0, a_{j}^\prime=1\} &= \{i_1^\prime<i_2^\prime<\ldots<i_N^\prime\}\label{eq2.38}\\
      \{j\in\mathbb{Z}_{\geq0}\mid a_{j-1}^\prime=1, a_{j}^\prime=0\} &= \{k_1^\prime<k_2^\prime<\ldots<k_N^\prime\}.\label{eq2.39}
     \end{align}
     From Eq. (\ref{eq2.28}), we have $i_m^\prime=i_m, k_m^\prime=k_m(m=1, 2, \ldots, N)$.
      
     Letting $k_0=k_0^\prime=-1$, we have
     \begin{align}
      \tilde{b}_j &= \begin{cases}
       \tilde{\eta}_j & k_m+1\leq j\leq i_{m+1}-1,\\
       \tilde{L}-\tilde{\eta}_{j+1} & i_m\leq j\leq k_{m}-2,\\
       0 & j=k_m-1, k_m.
      \end{cases}\quad(m=0,1,2,\ldots)\label{eq2.40}
     \end{align}
     from Eq. (\ref{eq2.23}), and
     \begin{align}
      \tilde{b}_j^\prime &= \begin{cases}
       \tilde{\eta}_j & k_m^\prime+1\leq j\leq i_{m+1}^\prime-1,\\
       \tilde{L}-\tilde{\eta}_{j-1} & i_m^\prime+2\leq j\leq k_{m}^\prime,\\
       0 & j=i_m^\prime, i_m^\prime+1.
      \end{cases}\quad(m=0,1,2,\ldots)\label{eq2.41}
     \end{align}
     from Eq. (\ref{eq2.30}).
     We obtain $\tilde{\BS{c}}$ from $\tilde{\BS{b}}$ by skipping the terms $j=k_m-1, k_m$, and $\tilde{\BS{c}}^\prime$ from $\tilde{\BS{b}}^\prime$ by skipping the terms $j=i_m^\prime, i_m^\prime+1$. Thus, we obtain $\tilde{\BS{c}}=\tilde{\BS{c}}^\prime$ and $\BS{c}=\BS{c}^\prime$.
     
     Letting $\tilde{\eta}_{-1}=\tilde{L}$, we can write the soliton sequence $\tilde{\BS{d}}$ as $\tilde{\BS{d}}=0^{\tilde{s}_1}1^{\tilde{t}_1}0^{\tilde{s}_2}1^{\tilde{t}_2}\cdots1^{\tilde{t}_N}0^\infty$, where 
     \begin{align}
      \tilde{s}_m &= \sum_{j=k_{m-1}}^{i_m} (\tilde{L}-\tilde{\eta}_j) - \sum_{j=k_{m-1}+1}^{i_{m}-1} \tilde{b}_j,\label{eq2.42}\\
      \tilde{t}_m  &= \sum_{j=i_m}^{k_m} \tilde{\eta}_j - \sum_{j=i_m}^{k_m-1} \tilde{b}_j, \label{eq2.43}
     \end{align}
     and $\BS{d}^\prime$ as $\BS{d}^\prime=0^{\tilde{s}_1^\prime}1^{\tilde{t}_1^\prime}0^{\tilde{s}_2^\prime}1^{\tilde{t}_2^\prime}\cdots1^{\tilde{t}_N^\prime}0^\infty$, where 
     \begin{align}
      \tilde{s}_1^\prime &= \sum_{j=0}^{i_1^\prime} (\tilde{L}-\tilde{\eta}_j) - \sum_{j=0}^{i_{1}^\prime-1} \tilde{b}_j^\prime+M,\label{eq2.44}\\
      \tilde{s}_m^\prime &= \sum_{j=k_{m-1}^\prime}^{i_m^\prime} (\tilde{L}-\tilde{\eta}_j) - \sum_{j=k_{m-1}^\prime+1}^{i_{m}^\prime-1} \tilde{b}_j^\prime\quad (m=2, 3, \ldots, N),\label{eq2.45}\\
      \tilde{t}_m^\prime  &= \sum_{j=i_m^\prime}^{k_m^\prime} \tilde{\eta}_j - \sum_{j=i_m^\prime+1}^{k_m^\prime} \tilde{b}_j^\prime\quad (m=1, 2, \ldots, N)\label{eq2.46}.
     \end{align}
     Using the Eqs. (\ref{eq2.40}), (\ref{eq2.41}) and $i_m^\prime=i_m, k_m^\prime=k_m$,  we have
     \begin{align}
      \tilde{s}_1 = \tilde{s}_1^\prime-M &= \tilde{L} + \sum_{j=-1}^{i_1-1}(\tilde{L}-\tilde{\eta}_j-\tilde{\eta}_{j+1}),\label{eq2.47}\\
      \tilde{s}_m = \tilde{s}_m^\prime &= \tilde{L} + \sum_{j=k_{m-1}}^{i_m-1}(\tilde{L}-\tilde{\eta}_j-\tilde{\eta}_{j+1})\quad(m=2, 3, \ldots, N),\label{eq2.48}\\
      \tilde{t}_m = \tilde{t}_m^\prime &= \tilde{L} + \sum_{j=i_m}^{k_m-1}(\tilde{\eta}_j+\tilde{\eta}_{j+1}-\tilde{L})\quad(m=1, 2, \ldots, N)\label{eq2.49}.
     \end{align}
     Therefore, we get $\tilde{\BS{d}}(\BS\eta)=\tilde{\BS{d}}^\prime(\BS\eta)$. Since $\tilde{L}-\tilde{\eta}_j-\tilde{\eta}_{j+1}=L-\eta_j-\eta_{j+1}$, the soliton sequences $\BS{d}=0^{s_1}1^{t_1}0^{s_2}1^{t_2}\cdots1^{t_N}0^{\infty}$ and $\BS{d}^\prime=0^{s_1^\prime}1^{t_1^\prime}0^{s_2^\prime}1^{t_2^\prime}\cdots1^{t_N^\prime}0^{\infty}$ can be written as
     \begin{align}
      s_m=s_m^\prime &= L+\sum_{j=k_{m-1}}^{i_{m}-1}(L-\eta_j-\eta_{j+1}),\label{eq2.50}\\
      t_m=t_m^\prime &= L+\sum_{j=i_m}^{k_m-1}(\eta_j+\eta_{j+1}-L)\quad (m=1, 2, \ldots, N).\label{eq2.51}
     \end{align}
    \end{proof}
   \end{theorem}

    From Theorems \ref{th2.9} and \ref{th2.12}, we obtain Eqs. (\ref{eq2.25}) and (\ref{eq2.26}) in Theorem \ref{th2.5} as
    \begin{align*}
     \Lambda(\BS{c}(\BS\eta)=\BS{c}^\prime(T_L(\BS\eta))=\BS{c}(T_L(\BS\eta)),\\
     T_1(\BS{d}(\BS\eta))=\BS{d}^\prime(T_L(\BS\eta))=\BS{d}(T_L(\BS\eta)).
    \end{align*}

    A binary sequence $\BS{d}\in\mathcal{S}_L^{(f)}$ is associated with a rigged configuration by the KKR bijection\cite{KKR}, and the time evolution of BBS with box capacity one can be linearized. We consider the case $\BS{d}\in\mathcal{S}_L^{(f)}\subset\mathcal{S}_1$, but this linearization property holds on $\mathcal{S}_1$.
    We briefly review the definition of a rigged configuration\cite{KOSTY}. Consider a partition $\mu=(\mu_1, \ldots, \mu_m)$. Define $m_j$ as the number of rows in $\mu$ whose lengths are $j$\,($j=1, \ldots, \mu_1$). A rigged configuration is a set $(\mu, J)$, where $J=(J_{1}, J_{2}, \ldots, J_{\mu_1})$, $J_{k}=(J_{k, 1}, J_{k, 2}, \ldots, J_{k, m_{k}})\in(\mathbb{Z})^{m_k}$. $J_{k, 1}, J_{k, 2}, \ldots, J_{k, m_k}$ are the {\it riggings} corresponding to the rows of length $k$.

   \begin{theorem}(Theorem 3 and Theorem 12 in Kakei et al. \cite{KNTW})\label{th2.13}

    Let $(\mu, J)$ be the rigged configuration associated with $\BS{d}\in\mathcal{S}_1$, and $(\overline{\mu}, \overline{J})$ be the rigged configuration associated with $T_1(\BS{d})$. Then, 
    \begin{align}
     \overline{\mu}_i &= \mu_i\quad(i=1, \ldots, m)\label{eq2.52}\\
     \overline{J}_{k, l} &= J_{k, l} + k\quad(k=1, \ldots, \mu_1, l=1, \ldots, m_k). \label{eq2.53}
    \end{align}
   \end{theorem}

   \begin{example}$L=2$\label{ex2.14}
    \begin{align*}
     \BS\eta &= 0, 0, 3, 1, -1, 2, 1, 1, 0, 0, \ldots\\
     \BS{c}(\BS\eta) &= 0, -1, -1, 1, 0, 0,  \ldots\\
     \BS{d}(\BS\eta) &= 0000111110000011100000\cdots\\
     T_2(\BS\eta) &= 0, 0, -1, 1, 3, 0, 1, 1, 2, 0, \ldots\\
     \BS{c}(T_2(\BS\eta)) &= 0, 0, -1, -1, 1, 0, \ldots\\
     \BS{d}(T_2(\BS\eta)) &= 00000000011111000111000000\cdots
    \end{align*}
    \begin{figure}[H]
     \centering
     \scalebox{0.9}{\begin{tikzpicture}[auto, ->]
      \node (a) at (0, 2) {$\BS\eta$};
      \node (x) at (7, 2) {$T_2(\BS\eta)$};
      \node (b) at (0, 0) {$\left\{\begin{array}{l}\BS{c}(\BS\eta),\\\BS{d}(\BS\eta)\end{array}\right\}$};
      \node (y) at (7, 0) {$\left\{\begin{array}{ll}\BS{c}(T_2(\BS\eta)),\\\BS{d}(T_2(\BS\eta))\end{array}\right\}$};
      \newcommand\mone{-1}
      \node (c) at (0, -3) {$\left\{\begin{array}{l}\BS{c}(\BS\eta),\\ \gyoung(;;;;;:\mone,;;;:5)\end{array}\right.$};
      \node (z) at (7, -3) {$\left\{\begin{array}{l}\BS{c}(T_2(\BS\eta)),\\ \gyoung(;;;;;:4,;;;:8)\end{array}\right.$};
      \draw (a) -- node {$T_2$} (x);
      \draw (x) -- node {$\beta_2$} (y);
      \draw (a) -- node[swap] {$\beta_2$} (b);
      \draw (b) -- node[swap, align=left] {$\BS{c}(T_2(\BS\eta))=\Lambda(\BS{c}(\BS\eta))$\\ $\BS{d}(T_2(\BS\eta))=T_1(\BS{d}(\BS\eta))$} (y);
      \draw (b) -- node[swap] {} (c);
      \draw (y) -- node[swap] {} (z);
      \draw (c) -- node {} (z);
     \end{tikzpicture}}
    \end{figure}
   \end{example}

  %%%%%%%%%%%%%%%%%%%%%%%%%%%%%   RECONSTRUCTION
    
  \section{\texorpdfstring{Reconstruction of BBS Sequence $\BS\eta$ and Bijectivity of $\beta_L$}{}} \label{sec3}
   In this section, we prove the bijectivity of the map $\beta_L$. First, we define the reconstruction map $\beta_L^{-1}: \mathcal{S}_{L, M}^{(b)}\oplus\mathcal{S}_L^{(f)}\to\mathcal{S}_{L, M}; (\BS{c}, \BS{d})\mapsto\BS\eta$ as follows.

    Let $M=\max_{j\in\mathbb{Z}_{\geqq0}}\max(-c_j, c_j-L), \tilde{L}=L+2M$. Let $\tilde{\BS{c}}=(\tilde{c}_0, \tilde{c}_1, \ldots)$ where $\tilde{c}_j=c_j+M$, and $\tilde{c}_{-1}=0$. When $\BS{d}$ is represented as $\BS{d}=0^{s_1}1^{t_1}\cdots1^{t_N}0^\infty$, let $\tilde{\BS{d}}=0^{s_1+M}1^{t_1+2M}0^{s_2+2M}1^{t_2+2M}\cdots1^{t_N+2M}0^\infty$.

    In the algorithm below, we will define variables $k^{(j)}$ and sequences $\BS{\tilde{c}}^{(j)}$ for $j=0, 1, 2, \ldots$. The integer $k^{(j)}$ will denote the position in the subsequence of $\tilde{d}$ that corresponds to the left end of the $j$-th box of $\BS\eta$ (or $\BS{\tilde\eta}$), and let $k^{(0)}=0$. The sequence $\tilde{\BS{c}}^{(j)}$ will be obtained by inserting 0s into $\tilde{\BS{c}}$ up to the $j$-th box of $\BS\eta$, with $\tilde{\BS{c}}^{(0)}=\tilde{\BS{c}}$. Repeat the following procedure for $j=0, 1, 2, \ldots$ in this order, and stop if $\tilde{c}_l^{(j)}=M$ for all $l > j$ and $\tilde{d}_m=0$ for all $m>k^{(j)}$. 
    \begin{enumerate}
     \renewcommand{\labelenumi}{\arabic{enumi}).}
     \item For the binary sequence $\tilde{\BS{d}}$, let $X_j=i-k^{(j)}$ where $i$ is the minimal integer that satisfies $i\geq k^{(j)}$, $\tilde{d}_{i-1}=1$, and $\tilde{d}_{i}=0$. If there is no such integer $i$, let $X_j=+\infty$.
     \item
      \begin{itemize}
       \item Case 1: $X_j<2\tilde{L}-\tilde{c}_{j-1}^{(j)}-\tilde{c}_j^{(j)}$

        Let $\tilde{\BS{c}}^{(j+1)}$ be the sequence obtained by inserting 0 between $\tilde{c}_{j-1}^{(j)}$ and $\tilde{c}_{j}^{(j)}$. The length of the subsequence of $\tilde{\BS{d}}$ that corresponds to the $j$-th box of $\BS\eta$ is $\tilde{L}-\tilde{c}_{j-1}^{(j+1)}-\tilde{c}_{j}^{(j+1)}=\tilde{L}-\tilde{c}_{j-1}^{(j+1)}$ and let $k^{(j+1)}$ be
        \begin{align*}
         k^{(j+1)}&=k^{(j)}+\left(\tilde{L}-\tilde{c}_{j-1}^{(j+1)}\right).
        \end{align*}
        Underline the subsequence of $\tilde{\BS{d}}$, from $\tilde{d}_{k^{(j)}}$ to $\tilde{d}_{k^{(j+1)}-1}$.
    
       \item Case 2: $X_j\geq2\tilde{L}-\tilde{c}_{j-1}^{(j)}-\tilde{c}_j^{(j)}$
     
        Let $\tilde{\BS{c}}^{(j+1)}=\tilde{\BS{c}}^{(j)}$. Then, the length of the subsequence of $\tilde{\BS{d}}$ that corresponds to the $j$-th box of $\BS\eta$ is $\tilde{L}-\tilde{c}_{j-1}^{(j+1)}-\tilde{c}_{j}^{(j+1)}$, and let $k^{(j+1)}$ be
        \begin{align*}
         k^{(j+1)} &= k^{(j)} + \left(\tilde{L}-\tilde{c}_{j-1}^{(j+1)}-\tilde{c}_{j}^{(j+1)}\right).
        \end{align*}
        Underline the subsequence of $\tilde{\BS{d}}$, from $\tilde{d}_{k^{(j)}}$ to $\tilde{d}_{k^{(j+1)}-1}$.
      \end{itemize}
    \end{enumerate}
    After this procedure stops at step $j=J$, $\tilde{\eta}_i$ is obtained by summing $\tilde{c}_i^{(J)}$ with the number of 1s over the $j$-th underline in $\tilde{d}$ if $i = 0, 1, \ldots, J-1$. For $i\geq J$, $\tilde{\eta}_i = \tilde{c}_i^{(J)}$. 

   \begin{example}\label{ex3.1}
    Consider $L = 2$, $\BS{c}=0, -1, -1, 1, 0, \ldots$, $\BS{d}=00001111100000111000\cdots$.
    Then, $M=1$, $\tilde{L}=4$, $\tilde{\BS{c}}=1, 0, 0, 2, 1, \ldots$, $\tilde{\BS{d}}=00000111111100000001111100000\cdots$.
    \begin{itemize}
     \item($j=0$) $k^{(0)}=0$, $i=12$, and $X_0=12$.

      $2\tilde{L}-\tilde{c}_{-1}^{(0)}-\tilde{c}_{0}^{(0)} = 7 \leq 12$

      $\tilde{\BS{c}}^{(1)} = 1, 0, 0, 2, 1, \ldots$, $\tilde{L}-\tilde{c}_{-1}^{(1)}-\tilde{c}_0^{(1)}=3$.

      $\tilde{\BS{d}}=\underline{000}00111111100000001111100000\cdots$
     \item($j=1$) $k^{(1)}=3$, $i=12$, and $X_1=9$.

      $2\tilde{L}-\tilde{c}_{0}^{(1)}-\tilde{c}_{1}^{(1)} = 7 \leq 9$

      $\tilde{\BS{c}}^{(2)} = 1, 0, 0, 2, 1, \ldots$, $\tilde{L}-\tilde{c}_{0}^{(2)}-\tilde{c}_1^{(2)}=3$.

      $\tilde{\BS{d}}=\underline{000}\,\underline{001}11111100000001111100000\cdots$
     \item($j=2$) $k^{(2)}=6$, $i=12$, and $X_2=6$.

      $2\tilde{L}-\tilde{c}_{1}^{(2)}-\tilde{c}_{2}^{(2)} = 8 > 6$

      $\tilde{\BS{c}}^{(3)} = 1, 0, 0, 0, 2, 1, \ldots$, $\tilde{L}-\tilde{c}_{1}^{(3)}-\tilde{c}_2^{(3)}=4$.

      $\tilde{\BS{d}}=\underline{000}\,\underline{001}\,\underline{1111}1100000001111100000\cdots$
     \item($j=3$) $k^{(3)}=10$, $i=12$, and $X_3=2$.

      $2\tilde{L}-\tilde{c}_{2}^{(3)}-\tilde{c}_{3}^{(3)} = 8 > 2$

      $\tilde{\BS{c}}^{(4)} = 1, 0, 0, 0, 0, 2, 1, \ldots$, $\tilde{L}-\tilde{c}_{2}^{(4)}-\tilde{c}_3^{(4)}=4$.

      $\tilde{\BS{d}}=\underline{000}\,\underline{001}\,\underline{1111}\,\underline{1100}000001111100000\cdots$
     \item($j=4$) $k^{(4)}=14$, $i=24$, and $X_4=10$.

      $2\tilde{L}-\tilde{c}_{3}^{(4)}-\tilde{c}_{4}^{(4)} = 8 \leq 10$

      $\tilde{\BS{c}}^{(5)} = 1, 0, 0, 0, 0, 2, 1, \ldots$, $\tilde{L}-\tilde{c}_{3}^{(5)}-\tilde{c}_4^{(5)}=4$.

      $\tilde{\BS{d}}=\underline{000}\,\underline{001}\,\underline{1111}\,\underline{1100}\,\underline{0000}01111100000\cdots$
     \item($j=5$) $k^{(5)}=18$, $i=24$, and $X_5=6$.

      $2\tilde{L}-\tilde{c}_{4}^{(5)}-\tilde{c}_{5}^{(5)} = 6 \leq 6$

      $\tilde{\BS{c}}^{(6)} = 1, 0, 0, 0, 0, 2, 1, \ldots$, $\tilde{L}-\tilde{c}_{4}^{(6)}-\tilde{c}_5^{(6)}=2$.

      $\tilde{\BS{d}}=\underline{000}\,\underline{001}\,\underline{1111}\,\underline{1100}\,\underline{0000}\,\underline{01}111100000\cdots$
     \item($j=6$) $k^{(6)}=20$, $i=24$, and $X_6=4$.

      $2\tilde{L}-\tilde{c}_{5}^{(6)}-\tilde{c}_{6}^{(6)} = 5 > 4$

      $\tilde{\BS{c}}^{(7)} = 1, 0, 0, 0, 0, 2, 0, 1, \ldots$, $\tilde{L}-\tilde{c}_{5}^{(7)}-\tilde{c}_6^{(7)}=2$.

      $\tilde{\BS{d}}=\underline{000}\,\underline{001}\,\underline{1111}\,\underline{1100}\,\underline{0000}\,\underline{01}\,\underline{11}1100000\cdots$
     \item($j=7$) $k^{(7)}=22$, $i=24$, and $X_7=2$.

      $2\tilde{L}-\tilde{c}_{6}^{(7)}-\tilde{c}_{7}^{(7)} = 7 > 2$

      $\tilde{\BS{c}}^{(8)} = 1, 0, 0, 0, 0, 2, 0, 0, 1, \ldots$, $\tilde{L}-\tilde{c}_{6}^{(8)}-\tilde{c}_7^{(8)}=4$.

      $\tilde{\BS{d}}=\underline{000}\,\underline{001}\,\underline{1111}\,\underline{1100}\,\underline{0000}\,\underline{01}\,\underline{11}\,\underline{1100}000\cdots$
     \item($j=8$) $k^{(8)}=26$, $i=+\infty$, and $X_8=+\infty$.

      $2\tilde{L}-\tilde{c}_{7}^{(8)}-\tilde{c}_{8}^{(8)} = 7 \leqq +\infty$

      $\tilde{\BS{c}}^{(9)} = 1, 0, 0, 0, 0, 2, 0, 0, 1, \ldots$, $\tilde{L}-\tilde{c}_{7}^{(9)}-\tilde{c}_8^{(9)}=3$.

      $\tilde{\BS{d}}=\underline{000}\,\underline{001}\,\underline{1111}\,\underline{1100}\,\underline{0000}\,\underline{01}\,\underline{11}\,\underline{1100}\,\underline{000}\cdots$
    \end{itemize}
    Then, we obtain $\tilde{\BS\eta} = 1, 1, 4, 2, 0, 3, 2, 2, 1, \ldots$ and $\BS\eta=0, 0, 3, 1, -1, 2, 1, 1, 0, \ldots$.
   \end{example}

   \begin{theorem}\label{th3.2}(Injectivity of $\beta_L$)\\
    For $\BS\eta\in\mathcal{S}_{L, M}$, $(\beta_L^{-1}\circ\beta_L)(\BS\eta)=\BS\eta$.
    \begin{proof}
   If we have the background sequence $\tilde{\BS{b}}=(\tilde{b}_0, \tilde{b}_1, \ldots)\in\mathcal{S}_{L, M}^{(b)}$ and a soliton sequence $\tilde{\BS{d}}$, we can get a binary subsequence that corresponds to the $j$-th box of $\BS\eta$ by dividing $\tilde{\BS{d}}$ every $\tilde{L}-\tilde{b}_{j-1}-\tilde{b}_j$. Further, $\tilde{\eta}_j$ is the sum of $\tilde{b}_j$ and the number of 1s in the $j$-th subsequence. Let $\tilde{\BS{c}}$ and $\tilde{\BS{d}}$ respectively be the raised background sequence and soliton sequence constructed from $\BS\eta$. 
 Thus it is sufficient to show that $\tilde{\BS{b}}$ is uniquely determined from $\tilde{\BS{c}}$ and $\tilde{\BS{d}}$. In the following,
we will prove that $X_j < 2\tilde{L}-\tilde{c}_{j-1}^{(j)}-\tilde{c}_j^{(j)}$ if and only if $(a_i, a_{i+1})=(1, 0)$ or $(a_{i+1}, a_{i+2})=(1, 0)$.

   \begin{itemize}
    \item Case 1: $a_{i+2k-1}=1, a_{i+2k}=0$ ($k=1, 2, \ldots, n$) and $(a_{i+2n+1}, a_{i+2n+2})\neq (1, 0)$. (See Fig. \ref{fig6})

       \begin{align*}
        X_i &= (\tilde{L}-\tilde{\eta}_i-\tilde{c}_{i-1})+\tilde{\eta}_i+\tilde{\eta}_{i+1}\\
        &= \tilde{L} - \tilde{c}_{i-1}+\tilde{\eta}_{i+1}\\
        X_{i+1} &= \tilde{\eta}_{i+1}
       \end{align*}
       From $a_{i+2k-1}=1, a_{i+2k}=0$ ($k=1, 2, \ldots, n$), we get $\tilde{\eta}_{i+2k-2}+\tilde{\eta}_{i+2k-1}>\tilde{L}, \tilde{\eta}_{i+2k-1}+\tilde{\eta}_{i+2k}<\tilde{L}$ and $\tilde{b}_{i+2n}=\tilde{c}_i\leq \tilde{\eta}_{i+2n}$. Therefore
       \begin{align*}
        X_i &< (\tilde{L}-\tilde{c}_{i-1}+\tilde{\eta}_{i+1})+\sum_{k=1}^{n}\left(\tilde{L}-\tilde{\eta}_{i+2k-1}-\tilde{\eta}_{i+2k}\right)\\
        &\quad\quad+\sum_{k=2}^{n}\left(\tilde{\eta}_{i+2k-2}+\tilde{\eta}_{i+2k-1}-\tilde{L}\right)\\
        &= 2\tilde{L}-\tilde{c}_{i-1}-\tilde{\eta}_{i+2n}\\
        &\leq 2\tilde{L}-\tilde{c}_{i-1}-\tilde{c}_i, \\
        X_{i+1} &= X_i-(\tilde{L}-\tilde{c}_{i-1})\\
         &< \tilde{L}-\tilde{c}_i\\
         &\leq 2\tilde{L}-\tilde{c}_i-\tilde{c}_{i+1}.
       \end{align*}

    \item Case 2: $\tilde{b}_i=\tilde{c}_i=\tilde{L}-\tilde{\eta}_{i+1}$, $a_{i+1}=\cdots=a_{k}=1, a_{k+1}=0$. (We need not insert zeros. See Fig. \ref{fig7})

     \begin{align*}
      X_i &= (\tilde{L}-\tilde{\eta}_i-\tilde{c}_{i-1})+\sum_{j=i}^{k-2} (\tilde{\eta}_{j}+\tilde{\eta}_{j+1}-\tilde{L})+\tilde{\eta}_{k-1}+\tilde{\eta}_{k}\\
       &= \tilde{L}-\tilde{c}_{i-1}+\tilde{\eta}_{i+1}+\sum_{j=i+1}^{k-1}(\tilde{\eta}_{j}+\tilde{\eta}_{j+1}-\tilde{L})\\
     \end{align*}
     From $a_{j+1}=1$, we get $\tilde{\eta}_j+\tilde{\eta}_{j+1}-\tilde{L}\geq 0$ for $j=i+1, \ldots, k-1$, and from $\tilde{b}_i=\tilde{c}_i=\tilde{L}-\tilde{\eta}_{i+1}$, we get $\tilde{\eta}_{i+1}=\tilde{L}-\tilde{c}_{i}$, 
     \begin{align*}
      X_i &\geq \tilde{L} + (\tilde{L} - \tilde{c}_i) - \tilde{c}_{i-1}\\
       &= 2\tilde{L}-\tilde{c}_{i-1}-\tilde{c}_i
     \end{align*}

    \item Case 3: $\tilde{b}_i=\tilde{c}_i=\tilde{\eta}_i$, $\tilde{a}_i=\cdots=\tilde{a}_l=0, \tilde{a}_{l+1}=\cdots=\tilde{a}_{k}=1$. (We need not insert zeros. See Fig. \ref{fig8})

     \begin{align*}
      X_i &= (\tilde{L}-\tilde{\eta}_i-\tilde{c}_{i-1})+\sum_{j=i}^{l-1}(\tilde{L}-\tilde{\eta}_j-\tilde{\eta}_{j+1})\\
      &\quad\quad + \sum_{j=l}^{k-2}(\tilde{\eta}_{j}+\tilde{\eta}_{j+1}-\tilde{L})+\eta_{k-1}+\eta_k\\
       &= 2\tilde{L}-\tilde{\eta}_i-\tilde{c}_{i-1}+\sum_{j=i}^{l-1}(\tilde{L}-\tilde{\eta}_j-\tilde{\eta}_{j+1})\\
       &\quad\quad + \sum_{j=l}^{k-1}(\tilde{\eta}_{j}+\tilde{\eta}_{j+1}-L)
     \end{align*}
     For $j=i, \ldots,l-1$, from $a_{j+1}=0$, we get $\tilde{L}-\tilde{\eta}_j-\tilde{\eta}_{j+1}\geq 0$, and for $j=l, \ldots, k-1$, from $a_{j+1}=1$, we get $\tilde{\eta}_j+\tilde{\eta}_{j+1}-\tilde{L}\geq0$. Recalling $\tilde{\eta}_i=\tilde{c}_i$, 
     \begin{align*}
      X_i \geq 2\tilde{L}-\tilde{c}_{i-1}-\tilde{c}_i.
     \end{align*}

     \begin{figure*}[htbp]\centering
        % \begin{tabular}{cccccccc}
        %  $\tilde{\BS{\eta}}:$ &  $\tilde{\eta}_i$ & $\tilde{\eta}_{i+1}$ & $\tilde{\eta}_{i+2}$ & $\cdots$ & $\tilde{\eta}_{i+2n-1}$ & $\tilde{\eta}_{i+2n}$ & $\cdots$\\
        %  $\BS{a}:$ &  $a_{i}$ & $a_{i+1}=1$ & $a_{i+2}=0$ & $\cdots$ & $a_{i+2n-1}=1$ & $a_{i+2n}=0$ & $\cdots$\\
        %  $\widetilde{\rbin}(\BS{\eta}):$ &  $0^{\tilde{L}-\tilde{\eta}_i}1^{\tilde{\eta}_i}$ & $1^{\tilde{\eta}_{i+1}}0^{\tilde{L}-\tilde{\eta}_{i+1}}$ & $0^{\tilde{L}-\tilde{\eta}_{i+2}}1^{\tilde{\eta}_{i+2}}$ & $\cdots$ & $1^{\tilde{\eta}_{i+2n-1}}0^{\tilde{L}-\tilde{\eta}_{i+2n-1}}$ & $0^{\tilde{L}-\tilde{\eta}_{i+2n}}1^{\tilde{\eta}_{i+2n}}$ & $\cdots$\\
        %  $\tilde{\BS{b}}:$ & 0 & 0 & 0 & $\cdots$ & 0 & $\tilde{b}_{i+2n}=\tilde{c}_i$ & $\cdots$\\
        %  $\tilde{\BS{c}}:$ & $\tilde{c}_i$ & $\tilde{c}_{i+1}$ & $\tilde{c}_{i+2}$ & $\cdots$ & $\cdots$ & $\cdots$ & $\cdots$\\
        %  $\BS{d}:$ &  $0^{\tilde{L}-\tilde{\eta}_i-\tilde{c}_{i-1}}1^{\tilde{\eta}_i}$ & $1^{\tilde{\eta}_{i+1}}0^{\tilde{L}-\tilde{\eta}_{i+1}}$ & $0^{\tilde{L}-\tilde{\eta}_{i+2}}1^{\tilde{\eta}_{i+2}}$ & $\cdots$ & $1^{\tilde{\eta}_{i+2n-1}}0^{\tilde{L}-\tilde{\eta}_{i+2n-1}}$ & $0^{\tilde{L}-\tilde{\eta}_{i+2n}}1^{\tilde{\eta}_{i+2n}-\tilde{c}_{i}}$ & $\cdots$
        % \end{tabular}
        \includegraphics[height=3.3cm]{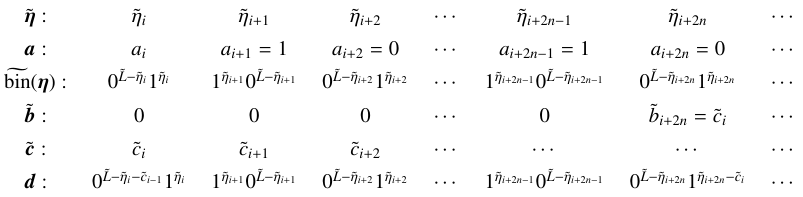}
        \caption{\label{fig6}The case (1) in Theorem \ref{th3.2}: $a_{i+2k-1}=1, a_{i+2k}=0$ ($k=1, 2, \ldots, n$) and $(a_{i+2n+1}, a_{i+2n+2})\neq (1, 0)$}\vspace{7pt}
      \end{figure*}

      \begin{figure*}[htbp]\centering
      % \begin{tabular}{cccccccc}
      %  $\tilde{\BS{\eta}}:$ & $\tilde{\eta}_i$ & $\tilde{\eta}_{i+1}$ & $\cdots$ & $\tilde{\eta}_{k-2}$ & $\tilde{\eta}_{k-1}$ & $\tilde{\eta}_{k}$ & $\tilde{\eta}_{k+1}$\\
      %  $\BS{a}:$ &  & 1 & $\cdots$ & 1 & 1 & 1 & 0\\
      %  $\widetilde{\rbin}(\BS{\eta}):$ & $0^{\tilde{L}-\tilde{\eta}_i}1^{\tilde{\eta}_i}$ & $0^{\tilde{L}-\tilde{\eta}_{i+1}}1^{\tilde{\eta}_{i+1}}$ & $\cdots$ & $0^{\tilde{L}-\tilde{\eta}_{k-2}}1^{\tilde{\eta}_{k-2}}$ & $0^{\tilde{L}-\tilde{\eta}_{k-1}}1^{\tilde{\eta}_{k-1}}$ & $1^{\tilde{\eta}_{k}}0^{\tilde{L}-\tilde{\eta}_{k}}$ & $0^{\tilde{L}-\tilde{\eta}_{k+1}}1^{\tilde{\eta}_{k+1}}$\\
      %  $\tilde{\BS{b}}:$ & $\tilde{c}_i=\tilde{L}-\tilde{\eta}_{i+1}$ & $\tilde{c}_{i+1}=\tilde{L}-\tilde{\eta}_{i+2}$ & $\cdots$ & $\tilde{c}_{k-2}=\tilde{L}-\tilde{\eta}_{k-1}$ & 0 & 0 & $\tilde{c}_{k-1}$\\
      %  $\tilde{\BS{c}}:$ & $\tilde{c}_i$ & $\tilde{c}_{i+1}$ & $\cdots$ & $\tilde{c}_{k-2}$ & $\tilde{c}_{k-1}$ & $\tilde{c}_k$ & \\
      %  $\BS{d}:$ & $0^{\tilde{L}-\tilde{\eta}_i-\tilde{c}_{i-1}}1^{\tilde{\eta}_i+\tilde{\eta}_{i+1}-\tilde{L}}$ & $1^{\tilde{\eta}_{i+1}+\tilde{\eta}_{i+2}-\tilde{L}}$ & $\cdots$ & $1^{\tilde{\eta}_{k-2}+\tilde{\eta}_{k-1}-\tilde{L}}$ & $1^{\tilde{\eta}_{k-1}}$ & $1^{\tilde{\eta}_{k}}0^{\tilde{L}-\tilde{\eta}_{k}}$ & $0^{\tilde{L}-\tilde{\eta}_{k+1}}1^{\tilde{\eta}_{k+1}-\tilde{c}_{k-1}}$
      % \end{tabular}
      \includegraphics[height=3.3cm]{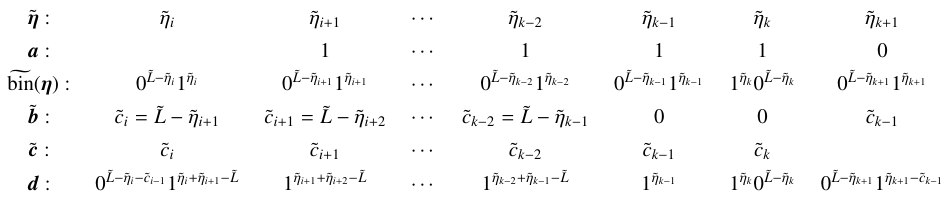}
      \caption{\label{fig7}The case (2) in Theorem \ref{th3.2}: $\tilde{b}_i=\tilde{c}_i=\tilde{L}-\tilde{\eta}_{i+1}$, $a_{i+1}=\cdots=a_{k}=1, a_{k+1}=0$}\vspace{7pt}
      \end{figure*}

      \begin{figure*}[htbp]\centering
      \includegraphics[height=3.3cm]{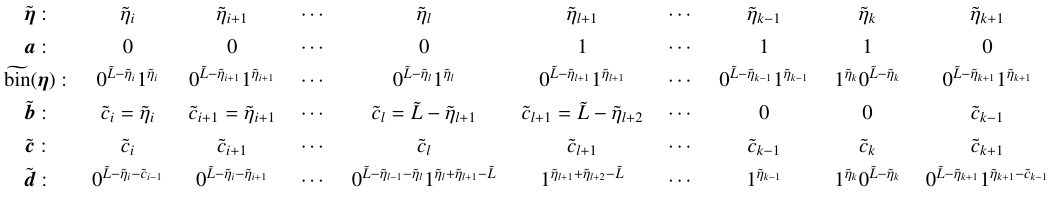}
      \caption{\label{fig8}The case (3) in Theorem \ref{th3.2}: $\tilde{b}_i=\tilde{c}_i=\tilde{\eta}_i$, $\tilde{a}_i=\cdots=\tilde{a}_l=0, \tilde{a}_{l+1}=\cdots=\tilde{a}_{k}=1$}
      \end{figure*}
   \end{itemize}
    \end{proof}
   \end{theorem}

   \begin{theorem}\label{th3.3}(Surjectivity of $\beta_L$)\\
    For $\BS{c}\in\mathcal{S}_{L, M}^{(b)}$, $\BS{d}\in\mathcal{S}_L^{(f)}$, $(\beta_L\circ\beta_L^{-1})(\BS{c}, \BS{d})=(\BS{c}, \BS{d})$.
    \begin{proof}
     For $\BS{c}\in\mathcal{S}_{L, M}^{(b)}$ and $\BS{d}\in\mathcal{S}_{L}^{(f)}$, let $\BS\eta=\beta_L^{-1}(\BS{c}, \BS{d})$, and let $\BS{a}$ be the soliton flag sequence calculated from $\BS\eta$. Let $Y_j$ be the number of consecutive 0s in soliton sequence $\BS{d}$ from $d_{k^{(j)}}$(If $d_{k^{(j)}}=1$, let $Y_j=0$). 
We will prove the following four statements: 
     \begin{enumerate}\renewcommand{\labelenumi}{(\arabic{enumi})}
      \item If $\tilde{L}\leq X_j<2\tilde{L}-\tilde{c}_{j-1}^{(j)}-\tilde{c}_j^{(j)}$, then $a_{j+1}=1$ and $a_{j+2}=0$.
      \item If $X_j<\tilde{L}$, then $a_j=1$ and $a_{j+1}=0$.
      \item If $X_j\geq2\tilde{L}-\tilde{c}_{j-1}^{(j)}-\tilde{c}_j^{(j)}$ and $Y_j<\tilde{L}-\tilde{c}_{j-1}^{(j)}-\tilde{c}_{j}^{(j)}$, then $a_{j+1}=a_{j+2}=1$.
      \item If $X_j\geq2\tilde{L}-\tilde{c}_{j-1}^{(j)}-\tilde{c}_j^{(j)}$ and $Y_j\geq\tilde{L}-\tilde{c}_{j-1}^{(j)}-\tilde{c}_{j}^{(j)}$, then $a_{j}=a_{j+1}=0$.
     \end{enumerate}
     In this proof, let $D_k$ be the subsequence of $\BS{d}$ corresponding to the $k$-th box of $\BS\eta$.
     \begin{enumerate}\renewcommand{\labelenumi}{(\arabic{enumi})}
      \item ($\tilde{L}\leq X_j<2\tilde{L}-\tilde{c}_{j-1}^{(j)}-\tilde{c}_{j}^{(j)}$. Shown in Fig. \ref{fig9}.)

     \begin{figure*}[htbp]\centering
      % \begin{tabular}{cc|ccc}
      %  $\tilde{\BS{c}}^{(j)}$: & $\tilde{c}_{j-1}^{(j)}$ & $\tilde{c}_j^{(j)}$ & & \\
      %  $\tilde{\BS{c}}^{(j+1)}$: & $\tilde{c}_{j-1}^{(j+1)}=\tilde{c}_{j-1}^{(j)}$ & $\tilde{c}_{j}^{(j+1)}=0$ & $\tilde{c}_{j+1}^{(j+1)}=\tilde{c}_{j}^{(j)}$ & \\
      %  $\tilde{\BS{c}}^{(j+2)}$: & $\tilde{c}_{j-1}^{(j+2)}=\tilde{c}_{j-1}^{(j)}$ & $\tilde{c}_{j}^{(j+2)}=0$ & $\tilde{c}_{j+1}^{(j+2)}=0$ & $\tilde{c}_{j+2}^{(j+2)}=\tilde{c}_{j}^{(j)}$\\
      %  $\tilde{\BS{d}}$: & & $\overbrace{0\cdots0\underbrace{1\cdots1}_{A}}^{\tilde{L}-\tilde{c}_{j-1}^{(j)}}$ & $\underbrace{1\cdots1}_{X_j-(\tilde{L}-\tilde{c}_{j-1}^{(j)})}0\cdots0$ & $\overbrace{0\cdots0\underbrace{1\cdots1}_{B}}^{\tilde{L}-\tilde{c}_{j}^{(j)}}$\\
      %  $\tilde{\BS\eta}$: & & $\tilde{\eta}_j=A+\tilde{c}_{j}^{(j+1)}$ & $\tilde{\eta}_{j+1}=X_j-\tilde{L}+\tilde{c}_{j-1}^{(j)}+\tilde{c}_{j+1}^{(j+2)}$ & $\tilde{\eta}_{j+2}=B+\tilde{c}_{j+2}^{(j+3)}$
      % \end{tabular}
      \includegraphics[height=3.7cm]{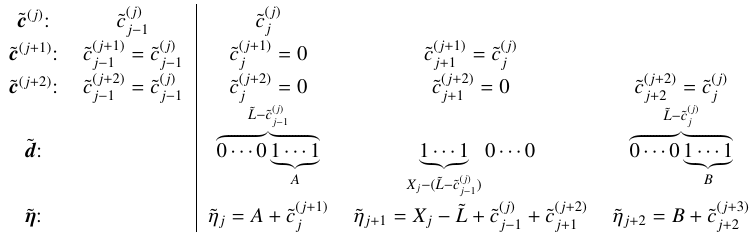}
      \caption{\label{fig9}The case (1) in Theorem \ref{th3.3}}
      \end{figure*}

      From the algorithm of reconstruction, we have $\tilde{c}_{j}^{(j+1)}=0$ and $\tilde{c}_{j-1}^{(j+1)}=\tilde{c}_{j-1}^{(j)}$. Since $X_j\geq \tilde{L}-\tilde{c}_{j-1}^{(j+1)}$, $X_{j+1}$ becomes
      \begin{align*}
       X_{j+1} &= X_j-\left(\tilde{L}-\tilde{c}_{j-1}^{(j+1)}\right)\\
        &< \left(2\tilde{L}-\tilde{c}_{j-1}^{(j)}-\tilde{c}_{j}^{(j)}\right)-\left(\tilde{L}-\tilde{c}_{j-1}^{(j)}\right)\\
        &=\tilde{L}-\tilde{c}_{j}^{(j)}\\
        &\leq \tilde{L}-\tilde{c}_{j}^{(j)}+\tilde{c}_{j}^{(j)}-\tilde{c}_{j}^{(j+1)}+\left(\tilde{L}-\tilde{c}_{j+1}^{(j+1)}\right)\\
        &= 2\tilde{L}-\tilde{c}_{j}^{(j+1)}-\tilde{c}_{j+1}^{(j+1)}, 
      \end{align*}
      and we get $\tilde{c}_{j+1}^{(j+2)}=0$.
      First, we prove $a_{j+1}=1$. Let $A$ denote the number of 1s in $D_j$.
      \begin{enumerate}\renewcommand{\labelenumii}{(\roman{enumii})}
       \item If $A<\tilde{L}-\tilde{c}_{j-1}^{(j)}$: 
        Since the number of consecutive 1s in $\BS{d}$ is larger than $\tilde{L}$, we get $A+X_j-(\tilde{L}-\tilde{c}_{j-1}^{(j)})>\tilde{L}$. Then, 
        \begin{align*}
         \tilde{\eta}_j+\tilde{\eta}_{j+1}-\tilde{L} &= A+X_j-\tilde{L}+\tilde{c}_{j-1}^{(j)}-\tilde{L}\\
         &>0, 
        \end{align*} and we get $a_{j+1}=1$.
       \item If $A=\tilde{L}-\tilde{c}_{j-1}^{(j)}$: 
        Let $i$ denote the maximum index less than $j$ such that the left end of consecutive 1s in $\BS{d}$ is at $D_i$. From (i), $a_{i+1}=1$. For $k=i+2, \ldots, j+1$, 
        \begin{align*}
         \tilde{\eta}_{k-1}+\tilde{\eta}_{k}-\tilde{L} &= (\tilde{L}-\tilde{c}_{k-2}^{(j)})+(\tilde{L}-\tilde{c}_{k-1}^{(j)})-\tilde{L}\\
          &= \tilde{L}-\tilde{c}_{k-2}^{(j)}-\tilde{c}_{k-1}^{(j)}\\
          &\geq0, 
        \end{align*}
        and we get $a_{k}=1$. 
      \end{enumerate}
      Next, we prove $a_{j+2}=0$. Let $B$ denote the number of 1s in $D_{j+2}$. 
      \begin{enumerate}\renewcommand{\labelenumii}{(\roman{enumii})}
       \item If $B>0$: 
        Since the number of consecutive 0s in $\BS{d}$ is larger than $\tilde{L}$, we get
        \begin{align*}
         \left(\tilde{L}-(X_j-\tilde{L}+\tilde{c}_{j-1}^{(j)})\right)&+(\tilde{L}-\tilde{c}_{j}^{(j)}-B)\\
          &\quad= 2\tilde{L}-X_j-B-\tilde{c}_{j-1}^{(j)}-\tilde{c}_{j}^{(j)}\\
         &\quad> 0.
        \end{align*}
        Here, using $\tilde{c}_{j+2}^{(j+3)}=\tilde{c}_{j}^{(j)}$ or 0, we get
        \begin{align*}
         \tilde{\eta}_{j+1}+\tilde{\eta}_{j+2}-\tilde{L} &= (X_j-\tilde{L}+\tilde{c}_{j-1}^{(j)})+(B+\tilde{c}_{j+2}^{(j+3)})-\tilde{L}\\
          &= -2\tilde{L}+X_j+B+\tilde{c}_{j-1}^{(j)}+\tilde{c}_{j+2}^{(j+3)}\\
          &< 0, 
        \end{align*}
        and $a_{j+2}=0$.
       \item If $B=0$: 
        \begin{align*}
         \tilde{\eta}_{j+1}+\tilde{\eta}_{j+2}-\tilde{L} &= (X_j-\tilde{L}+\tilde{c}_{j-1}^{(j)})+\tilde{c}_{j+2}^{(j+3)}-\tilde{L}\\
          &< (2\tilde{L}-\tilde{c}_{j-1}^{(j)}-\tilde{c}_{j}^{(j)})+\tilde{c}_{j-1}^{(j)}+\tilde{c}_{j+2}^{(j+3)}-2\tilde{L}\\
          &= \tilde{c}_{j+2}^{(j+3)}-\tilde{c}_{j}^{j}\\
          &\leq 0
        \end{align*}
        and $a_{j+2}=0$.
      \end{enumerate}

     \item ($X_j<\tilde{L}$. Shown in Fig. \ref{fig10}.)

     \begin{figure}[htbp]   \centering
      % \begin{tabular}{cc|c}
      %  $\tilde{\BS{c}}^{(j)}$: & $\tilde{c}_{j-1}^{(j)}$ & $\tilde{c}_j^{(j)}$\\
      %  $\tilde{\BS{c}}^{(j+1)}$: & $\tilde{c}_{j-1}^{(j+1)}=\tilde{c}_{j-1}^{(j)}$ & $\tilde{c}_{j}^{(j+1)}=0$\\
      %  $\tilde{\BS{d}}$: & $\overbrace{0\cdots0\underbrace{1\cdots1}_{A}}^{\tilde{L}-\tilde{c}_{j-2}^{(j)}-\tilde{c}_{j-1}^{(j)}}$ & $\underbrace{1\cdots1}_{X_j}\underbrace{0\cdots0}_{\tilde{L}-X_j}$\\
      %  $\tilde{\BS\eta}$: & $\tilde{\eta}_{j-1}=A+\tilde{c}_{j-1}^{(j)}$ & $\tilde{\eta}_{j}=X_j$
      % \end{tabular}
      \includegraphics[height=3.7cm]{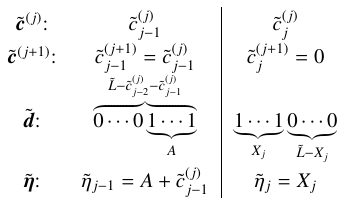}
      \caption{\label{fig10}The case (2) in Theorem \ref{th3.3}}
      \end{figure}

      If we assume that $X_{j-1}\geq 2\tilde{L}-\tilde{c}_{j-2}^{j-1}-\tilde{c}_{j-1}^{(j-1)}$ and $\tilde{c}_{j-1}^{(j)}=\tilde{c}_{j-1}^{(j-1)}$, then
      \begin{align*}
       X_{j-1} &= (\tilde{L}-\tilde{c}_{j-2}^{(j)}-\tilde{c}_{j-1}^{(j)})+X_j\\
        &< \tilde{L}-\tilde{c}_{j-2}^{(j-1)}-\tilde{c}_{j-1}^{(j-1)}+\tilde{L}\\
        &= 2\tilde{L}-\tilde{c}_{j-2}^{(j-1)}-\tilde{c}_{j-1}^{(j-1)}.
      \end{align*}
      Therefore, we get $X_{j-1}<2\tilde{L}-\tilde{c}_{j-2}^{j-1}-\tilde{c}_{j-1}^{(j-1)}$ and from Case (1) in Theorem \ref{th3.3}, $a_j=1$  and $a_{j+1}=0$.
      \item ($X_j\geq 2\tilde{L}-\tilde{c}_{j-1}^{(j)}-\tilde{c}_{j}^{(j)}$ and $Y_j<\tilde{L}-\tilde{c}_{j-1}^{(j)}-\tilde{c}_{j}^{(j)}$. Shown in Fig. \ref{fig11}.)

     \begin{figure}[htbp]   \centering
      % \begin{tabular}{cccc}
      %  $\tilde{\BS{c}}^{(j)}$: & $\tilde{c}_{j}^{(j)}$ & & \\
      %  $\tilde{\BS{c}}^{(j+1)}$: & $\tilde{c}_{j}^{(j+1)}=\tilde{c}_{j}^{(j)}$ & $\tilde{c}_{j+1}^{(j+1)}$ & \\
      %  $\tilde{\BS{c}}^{(j+2)}$: & $\tilde{c}_{j}^{(j+2)}=\tilde{c}_{j}^{(j)}$ & $\tilde{c}_{j+1}^{(j+2)}$ & $\tilde{c}_{j+2}^{(j+2)}$\\
      %  $\tilde{\BS{c}}^{(j+3)}$: & $\tilde{c}_{j}^{(j+3)}=\tilde{c}_{j}^{(j)}$ & $\tilde{c}_{j+1}^{(j+3)}=\tilde{c}_{j+1}^{(j+2)}$ & $\tilde{c}_{j+2}^{(j+3)}$\\
      %  $\tilde{\BS{d}}$: & $\overbrace{\underbrace{0\cdots0}_{Y_j}1\cdots1}^{\tilde{L}-\tilde{c}_{j-1}^{(j)}-\tilde{c}_{j}^{(j)}}$ & $\overbrace{1\cdots1}^{\tilde{L}-\tilde{c}_{j}^{(j)}-\tilde{c}_{j+1}^{(j+2)}}$ & $\overbrace{\underbrace{1\cdots1}_{B}0\cdots0}^{\tilde{L}-\tilde{c}_{j+1}^{(j+2)}-\tilde{c}_{j+2}^{(j+3)}}$\\
      %  $\tilde{\BS\eta}$: & $\tilde{\eta}_j=\tilde{L}-\tilde{c}_{j-1}^{(j)}-Y_j$ & $\tilde{\eta}_{j+1}=\tilde{L}-\tilde{c}_{j}^{(j)}$ & $\tilde{\eta}_{j+2}=B+\tilde{c}_{j+2}^{(j+3)}$
      % \end{tabular}
      \includegraphics[height=3.7cm]{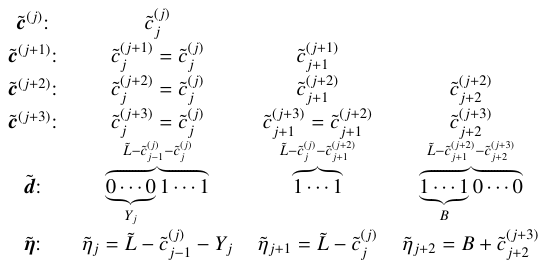}
      \caption{\label{fig11}The case (3) in Theorem \ref{th3.3}}
      \end{figure}
       First, 
       \begin{align*}
        \tilde{\eta}_{j}+\tilde{\eta}_{j+1}-\tilde{L} &= (\tilde{L}-\tilde{c}_{j-1}^{(j)}-Y_j)+(\tilde{\eta}_{j+1}=\tilde{L}-\tilde{c}_{j}^{(j)})-\tilde{L}\\
         &= \tilde{L}-\tilde{c}_{j-1}^{(j)}-\tilde{c}_{j}^{(j)}-Y_j\\
         &> 0, 
       \end{align*}
       and we get $a_{j+1}=1$.
       Next, we prove $a_{j+2}=1$. Let $B$ be the number of 1s in $D_{j+2}$.
      \begin{enumerate}\renewcommand{\labelenumii}{(\roman{enumii})}
       \item If $B<\tilde{L}-\tilde{c}_{j+1}^{(j+2)}-\tilde{c}_{j+2}^{(j+3)}$: Since 
        \begin{align*}
         B &= X_j-(\tilde{L}-\tilde{c}_{j-1}^{(j)}-\tilde{c}_{j}^{(j)})-(\tilde{L}-\tilde{c}_{j}^{(j)}-\tilde{c}_{j+1}^{(j+2)})\\
          &= \{X_j-(2\tilde{L}-\tilde{c}_{j-1}^{(j)}-\tilde{c}_{j}^{(j)})\}+\tilde{c}_{j}^{(j)}+\tilde{c}_{j+1}^{(j+2)}\\
          &\geq \tilde{c}_{j}^{(j)}+\tilde{c}_{j+1}^{(j+2)}, 
        \end{align*}
        we obtain
        \begin{align*}
         \tilde{\eta}_{j+1}+\tilde{\eta}_{j+2}-\tilde{L} &= B-\tilde{c}_{j}^{(j)}-\tilde{c}_{j+2}^{(j+3)}\\
          &\geq \tilde{c}_{j+1}^{(j+2)}+\tilde{c}_{j+2}^{(j+3)}\\
          &\geq 0,
        \end{align*}
        and then, $a_{j+2}=1$. 
       \item If $B=\tilde{L}-\tilde{c}_{j+1}^{(j+2)}-\tilde{c}_{j+2}^{(j+3)}$: 
       \begin{align*}
        \tilde{\eta}_{j+1}+\tilde{\eta}_{j+2}-\tilde{L} &= (\tilde{L}-\tilde{c}_{j}^{(j)})+(\tilde{L}-\tilde{c}_{j+1}^{(j+2)})-\tilde{L}\\
         &= \tilde{L}-\tilde{c}_{j}^{(j+2)}-\tilde{c}_{j+1}^{(j+2)}\\
         &\geq 0
       \end{align*}
       and $a_{j+2}=1$. 
      \end{enumerate}
      \item ($X_j\geq2\tilde{L}-\tilde{c}_{j-1}^{(j)}-\tilde{c}_j^{(j)}$ and $Y_j\geq\tilde{L}-\tilde{c}_{j-1}^{(j)}-\tilde{c}_{j}^{(j)}$. Shown in Fig. \ref{fig12}.)
     \begin{figure}[htbp]   \centering
      % \begin{tabular}{cc|cc}
      %  $\tilde{\BS{c}}^{(j)}$: & $\tilde{c}_{j-1}^{(j)}$ & $\tilde{c}_{j}^{(j)}$ & \\
      %  $\tilde{\BS{c}}^{(j+1)}$: & $\tilde{c}_{j-1}^{(j+1)}=\tilde{c}_{j-1}^{(j)}$ & $\tilde{c}_{j}^{(j+1)}=\tilde{c}_{j}^{(j)}$ & $\tilde{c}_{j+1}^{(j+1)}$ \\
      %  $\tilde{\BS{c}}^{(j+2)}$: & $\tilde{c}_{j-1}^{(j+2)}=\tilde{c}_{j-1}^{(j)}$ & $\tilde{c}_{j}^{(j+2)}=\tilde{c}_{j}^{(j)}$ & $\tilde{c}_{j+1}^{(j+2)}$\\
      %  $\tilde{\BS{d}}$: & $\overbrace{\underbrace{1\cdots1}_{A}0\cdots0}^{\tilde{L}-\tilde{c}_{j-2}^{(j)}-\tilde{c}_{j-1}^{(j)}}$ & $\overbrace{0\cdots0}^{\tilde{L}-\tilde{c}_{j-1}^{(j)}-\tilde{c}_{j}^{(j)}}$ & $\overbrace{0\cdots0\underbrace{1\cdots1}_{B}}^{\tilde{L}-\tilde{c}_{j}^{(j)}-\tilde{c}_{j+1}^{(j+2)}}$ \\
      %  $\tilde{\BS\eta}$: & $\tilde{\eta}_{j-1}=A+\tilde{c}_{j-1}^{(j)}$ & $\tilde{\eta}_{j}=\tilde{c}_{j}^{(j)}$ & $\tilde{\eta}_{j+1}=B+\tilde{c}_{j+1}^{(j+2)}$
      % \end{tabular}
      \includegraphics[height=3.7cm]{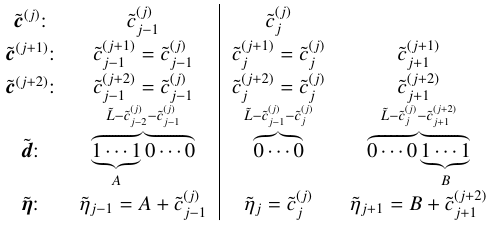}
       \caption{\label{fig12}The case (4) in Theorem \ref{th3.3}}
      \end{figure}

      First, we prove $a_j=0$.
      \begin{enumerate}\renewcommand{\labelenumii}{(\roman{enumii})}
       \item If $X_{j-1}\leq \tilde{L}-\tilde{c}_{j-2}^{(j)}-\tilde{c}_{j-1}^{(j)}$: 
        Since $X_{j-1}\leq\tilde{L}$ and Case (2) in Theorem \ref{th3.3}, we get $a_{j-1}=1, a_{j}=0$.
        \item If $X_{j-1}> \tilde{L}-\tilde{c}_{j-2}^{(j)}-\tilde{c}_{j-1}^{(j)}$: 
        Let $i$ denote  the maximum index less than $j$ such that the left end of consecutive 0s in $\BS{d}$ is at $D_i$. From (i), $a_i=1, a_{i+1}=0$. For $k=i+2, \ldots, j$, 
        \begin{align*}
         \tilde{\eta}_{k-1}+\tilde{\eta}_{k}-\tilde{L} &= \tilde{c}_{k-1}^{(j)}+\tilde{c}_{k}^{(j)}-\tilde{L}\\
          &\leq 0,
        \end{align*}
        and we get $a_{k}=0$.
      \end{enumerate}
      Next, we prove $a_{j+1}=0$. Let $B$ the number of 1s in $D_{j+1}$.
      \begin{enumerate}\renewcommand{\labelenumii}{(\roman{enumii})}
       \item If $B>0$: 
        Since $Y_j$ is the number of consecutive 0s in soliton sequence $\BS{d}$ from $d_{k^{(j)}}$, we have
        \begin{align*}
         Y_j + B = (\tilde{L}-\tilde{c}_{j-1}^{(j)}-\tilde{c}_{j}^{(j)})+(\tilde{L}-\tilde{c}_{j}^{(j)}-\tilde{c}_{j+1}^{(j+2)}).
        \end{align*}
        Using this, we obtain 
        \begin{align*}
         \tilde{\eta}_{j}+\tilde{\eta}_{j+1}-\tilde{L} &= B + \tilde{c}_{j}^{(j)}+\tilde{c}_{j+1}^{(j+2)}-\tilde{L}\\
          &= \tilde{L}-\tilde{c}_{j-1}^{(j)}-\tilde{c}_{j}^{(j)}-Y_j\\
          &\leq 0.
        \end{align*}
       \item If $B=0$: 
        \begin{align*}
         \tilde{\eta}_j+\tilde{\eta}_{j+1}-\tilde{L} &= \tilde{c}_{j}^{(j)}+\tilde{c}_{j+1}^{(j+2)}-\tilde{L}\\
          &\leq 0,
        \end{align*}
        and we obtain $a_j=a_{j-1}=0$.
      \end{enumerate}
     \end{enumerate}
    \end{proof}
   \end{theorem}
  
   From Theorems \ref{th3.2} and \ref{th3.3}, the bijectivity of the map $\beta_L$ is proved, and we have the following theorem.
   \begin{theorem}\label{th3.6}
    A BBS state with box capacity $L$ can be decomposed into a soliton sequence and a background sequence.
    \begin{align}
     \mathcal{S}_{L, M}=\mathcal{S}_{L, M}^{(b)}\otimes\mathcal{S}_{L}^{(f)}.\label{eq3.1}
    \end{align}
   \end{theorem}

 \section{Conclusion}
  We proposed a method to linearize the time evolution of BBS($L$) by decomposing a state into two sequences: a sequence that shifts to the right at speed 1 and a binary sequence that exhibits the time evolution of BBS($1$). For a state including a negative value or a value greater than the box capacity, this method is applicable with a simple variable transformation. 

 \section{Acknowledgement}
 %\begin{acknowledgment}
  The work of ST is partially supported by JSPS KAKENHI (Grant Numbers 19H01792). This research was partially supported by the joint project ``Advanced Mathematical Science for Mobility Society'' of Kyoto University and Toyota Motor Corporation.
 %\end{acknowledgment}

%\appendix
%\section{}

\end{document}